\documentclass[onecolumn,amsmath,amssymb,12pt,groupedaddress,nofootinbib]{revtex4-2}
\pdfoutput=1

\usepackage[latin1]{inputenc}
\usepackage[english]{babel}
\usepackage{amssymb}
\usepackage{amsmath}
\usepackage{amsthm}
\usepackage[]{graphicx}
\usepackage[]{subfigure}
\usepackage{tensor}
\usepackage{color}
\usepackage{cancel}
\usepackage{setspace}
\usepackage{fancyhdr}
\usepackage[bookmarks,linktocpage, colorlinks=true, plainpages = false, citecolor = blue,  linkcolor=blue, urlcolor = blue, filecolor = blue]{hyperref}

\begin{document}

\allowdisplaybreaks
\begin{titlepage}

\vspace{.2in}



\title{Lessons for quantum cosmology \\ from anti-de Sitter black holes
\vspace{.3in}}

\author{Alice Di Tucci}
\email{alice.di-tucci@aei.mpg.de}
\affiliation{Max Planck Institute for Gravitational Physics (Albert Einstein Institute),
14476 Potsdam, Germany}
\author{Michal P. Heller}
\email{mheller@aei.mpg.de}
\altaffiliation{\emph{On leave from:} National Centre for Nuclear Research, 02-093 Warsaw, Poland}
\affiliation{Max Planck Institute for Gravitational Physics (Albert Einstein Institute),
14476 Potsdam, Germany}
\author{Jean-Luc Lehners}
\email{jlehners@aei.mpg.de}
\affiliation{Max Planck Institute for Gravitational Physics (Albert Einstein Institute),
14476 Potsdam, Germany}

\begin{abstract}
\vspace{0.2in}
\noindent 
Gravitational physics is arguably better understood in the presence of a negative cosmological constant than a positive one, yet there exist strong technical similarities between the two settings. These similarities can be exploited to enhance our understanding of the more speculative realm of quantum cosmology, building on robust results regarding anti-de Sitter black holes describing the thermodynamics of holographic quantum field theories. To this end, we study $4$-dimensional gravitational path integrals in the presence of a negative cosmological constant, and with minisuperspace metrics. We put a special emphasis on boundary conditions and integration contours. The Hawking-Page transition is recovered and we find that below the minimum temperature required for the existence of black holes the corresponding saddle points become complex. When the asymptotic anti-de Sitter space is cut off at a finite distance, additional saddle points contribute to the partition function, albeit in a very suppressed manner. These findings have direct consequences for the no-boundary proposal in cosmology, because the anti-de Sitter calculation can be brought into one-to-one correspondence with a path integral for de Sitter space with Neumann conditions imposed at the nucleation of the universe. Our results lend support to recent implementations of the no-boundary proposal focusing on momentum conditions at the ``big bang''.
\end{abstract}
\maketitle

\end{titlepage}

\tableofcontents

\section{Introduction}

General relativity is a spectacularly successful theory of spacetime and gravity, but amongst its physically relevant solutions there are some that contain singularities and which thus predict the breakdown of the theory from which they originated. Most notable are the singularities inside black holes, and the big bang in cosmological solutions. A series of insights originating already in the 1970s and 1980s implied that black holes and the big bang, which may be seen as the most extreme manifestations of gravity in the universe, could be tamed when perceived from the point of view of Euclidean spacetime. In the case of black holes the Euclidean solution ends at the horizon, and the interior part containing the singularity is simply absent \cite{Hartle:1976tp}. For the big bang, the proposed resolution consists of rounding off the singularity by extending the spacetime to contain a non-singular Euclidean section near the big bang \cite{Hartle:1983ai}. 

The study of Euclidean black holes has been a fruitful way of deriving and elucidating thermodynamic properties of black holes, and thus also the link between gravity and quantum theory. In particular, Euclidean solutions offer the most pragmatic way of deriving the temperature of black holes. Another famous application for black holes in Anti-de Sitter (AdS) spacetime, on which we will focus here, is the Hawking-Page transition~\cite{Hawking:1982dh}. There one finds that, depending on the temperature, either empty AdS or black holes dominate the partition function while a minimum temperature is required for black hole solutions to exist at all. Via holography, gravitational physics in AdS acquires an alternative description in terms of more familiar quantum field theory (QFT) phenomena~\cite{Maldacena:1997re,Gubser:1998bc,Witten:1998qj}, which makes AdS the best understood instance of quantum gravity. In particular, the Hawking-Page transition in holography becomes a thermal phase transition between confined and deconfined phases in a dual QFT with the thermodynamic limit achieved by a very large number of underlying QFT degrees of freedom~\cite{Witten:1998zw}.

In this work we will reproduce features of the Hawking-Page phase transition by calculating explicitly the gravitational path integral in the minisuperspace approach~\cite{Halliwell:1988wc}. Our studies for black holes in AdS space are motivated by recent developments in cosmology as there is a very close analogy between the calculations performed in the present work and novel studies of the no-boundary proposal in cosmology utilising the minisuperspace ansatz~\cite{DiTucci:2019bui}. The idea is to make use of the firm results for the thermodynamics of black holes to learn lessons for path integrals in quantum cosmology. 

\begin{figure}[h]
\includegraphics[width=0.5\textwidth]{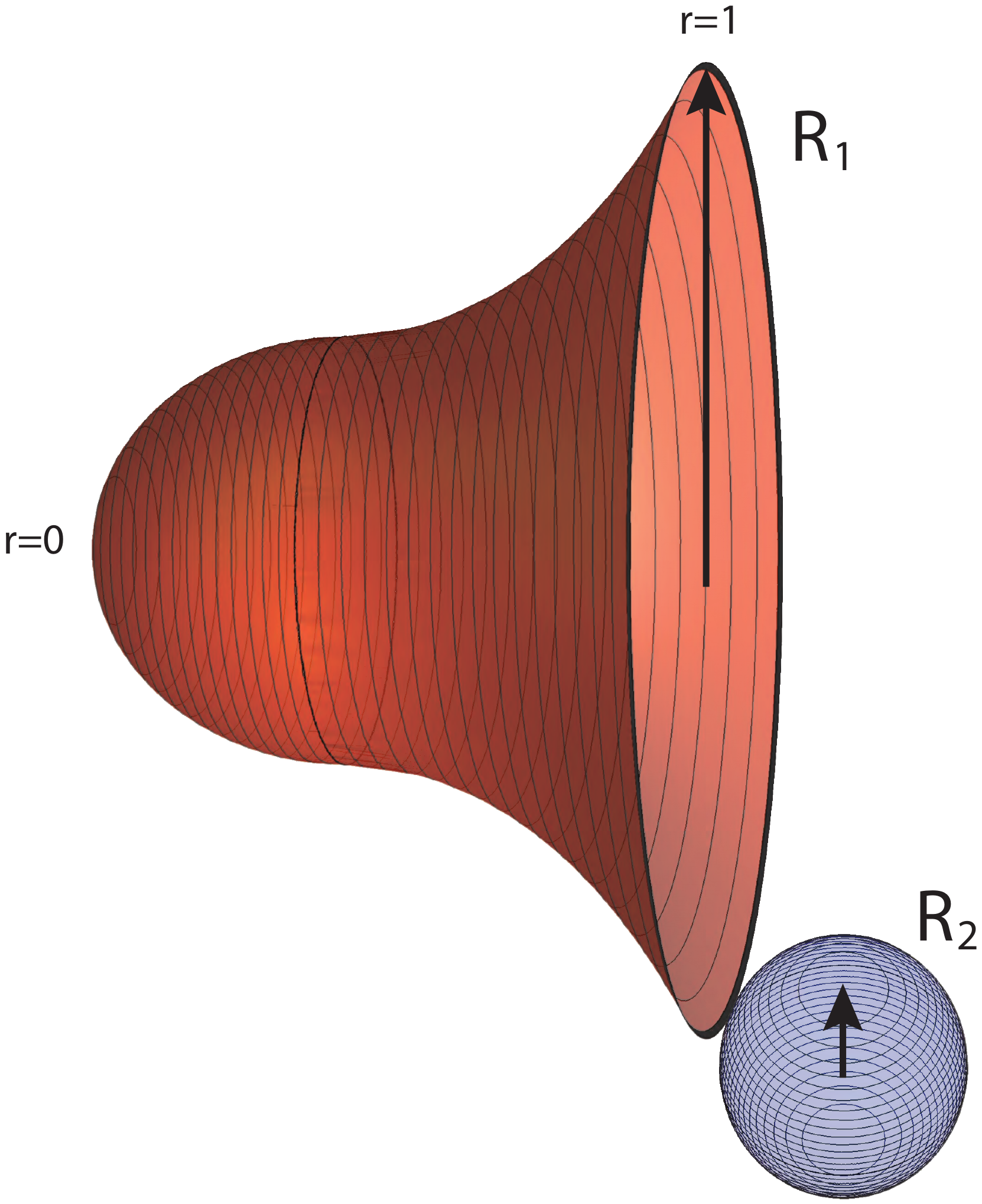}
\caption{We study partition functions that sum over all metrics with a fixed Euclidean boundary, where the boundary will either be a three-sphere or the product of a circle and a two-sphere. Illustrated here is the case with a fixed boundary consisting of the product of a circle of radius $R_1$ and a two-sphere of radius $R_2.$ The sphere is only shown at a single point on the circle. We will use coordinates in which the boundary resides at $r=1$ and we will sometimes refer to it as the ``outer'' boundary. We will assume that the geometry ends at $r=0$ and we will refer to the $r=0$ coordinate location as the ``inner'' boundary. Guided by holography, we will be interested in situations in which this location is not a true geometrical boundary but rather just the end point of a coordinate range.} \label{fig:partition}
\end{figure}

In the cosmological context one would also like to calculate path integrals which, instead of partition functions, are interpreted as describing transitions between an initial state of the universe and a final state which may usually be thought of as the current spatial section of the universe. The analogy is technically closest in the case where one models a possible early inflationary phase by a positive cosmological constant, and where the spatial section of the universe is taken to be a three-sphere. We will focus on this case here. This de~Sitter solution represents a useful approximation during an inflationary phase. The idea of Hartle and Hawking was that one could resolve the big bang singularity by gluing half of the Euclidean version of the solution to the waist of the Lorentzian de Sitter hyperboloid, so that the universe is smoothly rounded off in the past \cite{Hartle:1983ai}. This would replace the big bang by a semi-classical geometry and provide initial conditions for an inflationary phase in the early universe. 

A question that has been discussed for about forty years now is how to implement this proposal in a precise technical manner. Hartle and Hawking originally made the proposal that in the path integral one should sum only over compact, regular metrics. In this manner one should obtain a saddle point geometry corresponding to the rounded-off big bang they had in mind. This definition suggests using Dirichlet conditions in which the initial size of the universe is set to zero. However, one can show that if one implements a path integral with such Dirichlet conditions, then one necessarily obtains saddle points with unstable fluctuations \cite{Feldbrugge:2017kzv,Feldbrugge:2017fcc,Feldbrugge:2017mbc}. For such saddle points the weighting is larger when the fluctuations are larger, and this results in unphysical predictions\footnote{There were a number of papers that attempted to evade this conclusion by modifying integration contours or modifying boundary conditions for the perturbations, see~\cite{DiazDorronsoro:2017hti,DiazDorronsoro:2018wro,Vilenkin:2018dch,Vilenkin:2018oja,Feldbrugge:2018gin}.}. By contrast, path integrals with an initial Neumann condition were recently studied in the cosmological context in \cite{DiTucci:2019bui}\footnote{One may also consider more general conditions that form a linear combination of Dirichlet and Neumann~\cite{DiTucci:2019dji}. An early suggestion to use the Neumann condition can be found in \cite{Louko:1988bk, Halliwell:1988ik}.} and these were shown to allow for a consistent and stable formulation of the Hartle-Hawking wave function. 

In the present work we study gravitational path integrals in AdS with the aim of using holographic intuitions to shed light on the aforementioned studies in cosmology. To this end, we will consider path integrals over four-dimensional geometries within the minisuperspace class with weighting provided by the Einstein-Hilbert term + a negative cosmological constant + appropriate boundary terms. These geometries will be anchored on a Euclidean boundary and we will consider two separate cases of the latter: three-spheres and direct products of a circle and a two-sphere. According to holography, an appropriately understood gravitational path integral corresponds to evaluating the dual QFT partition function with the QFT living on the chosen boundary geometry. The latter we impose on the gravity side as a Dirichlet boundary condition for the four-dimensional metrics we path integrate over. Evaluating such path integrals in the minisuperspace approach requires us to introduce a coordinate~$r$ on which our metric will depend. Without loss of generality we can assume that this coordinate runs between $0$ and $1$ and we choose the Euclidean boundary to lie at $r = 1$. However, in order to make our calculation well-defined we will also need to provide information on how the metrics we integrate over behave at $r = 0$. To be more specific, for an ``outer'' boundary consisting of a circle of radius $R_1$ and a two-sphere of radius $R_2$ illustrated in Fig.~\ref{fig:partition} we will analyse the sum over geometries
\begin{align}
\label{eq.Zgravity}
Z(R_1,R_2) = \sum_{B} \int_{B}^{R_1,R_2} d[g_{\mu\nu}] e^{\frac{i}{\hbar}S}\,,
\end{align}
where $B$ encapsulates conditions imposed on the metric $g_{\mu \nu}$ at $r = 0$.

The first of our two main challenges will be to specify the boundary conditions~$B$ on the ``inner'' boundary, including possibly summing over a class of them. An important guiding principle for us in this quest will be the dual QFT interpretation. Following this thread, one thing that we do not want to do is to impose another Dirichlet boundary condition in which the locus $r = 0$ is a non-trivial three dimensional Euclidean space, since according to holography each such boundary corresponds to an independent copy of a QFT. As we will see later, interpreting the setup from Fig.~\ref{fig:partition} as calculating the thermodynamic partition function in a dual QFT implies that in our approach we must impose a Neumann condition on the metric at the inner boundary of the integration region at $r = 0$. Similarly, a Neumann boundary condition can also be utilised in the case of the boundary being a three-sphere, in which case the litmus test comes from a comparison with the exactly evaluated QFT partition function using supersymmetric localisation. Our AdS calculation in this case can be brought into one-to-one correspondence with the recent definition of the no-boundary proposal in \cite{DiTucci:2019bui} and shows that using momentum conditions at the big bang rather than summing over compact metrics is in fact quite natural from the holographic point of view.

The second main challenge in making sense of~\eqref{eq.Zgravity} will be evaluating the path integral itself. The appearance of meaningful Euclidean saddle point solutions, such as empty Euclidean AdS or a Euclidean AdS black hole, would naturally suggest that gravitational path integrals should be defined as sums over Euclidean geometries in the framework known as Euclidean Quantum Gravity~\cite{Gibbons:1976ue}. However, when using the minisuperspace path integral as a definition of~\eqref{eq.Zgravity}, with the exception of using a Neumann condition at $r = 0$ when the boundary is a three-sphere, we find that in general the sum over bulk Euclidean metrics is not given by a convergent integral and, therefore, is mathematically meaningless. Following~\cite{Feldbrugge:2017kzv}, we \emph{define} the path integral~\eqref{eq.Zgravity} as a sum over a class of sections of complex manifolds. This is necessary in order to turn the conditionally convergent integral \eqref{eq.Zgravity} into a sum of manifestly convergent integrals. These convergent integrals live on steepest descent contours (``Lefschetz thimbles'') and they fix both the meaning and the order of integration of the conditionally convergent sum over metrics. Our approach rests on the formalism of Picard-Lefschetz theory \cite{Witten:2010cx}, although we will only need the simplest, one-dimensional, version of the theory.

Let us already mention some consequences which we will explore. We find that in addition to well-known Euclidean saddle points describing empty AdS and AdS black holes, there always exist three other saddle points with Euclidean or complex bulk geometries. These play no role in the case where the outer boundary is sent off to infinity (i.e. $R_{1}$ and $R_{2}$ diverge with the ratio kept fixed) corresponding to dealing with a ultraviolet-complete holographic QFT [in our case, a holographic conformal field theory (CFT)]~\cite{Maldacena:1997re}. However, they do contribute, though in a very suppressed manner, to the path integral~\eqref{eq.Zgravity} when the boundary is brought to a finite radius (i.e. both $R_{1}$ and $R_{2}$ are finite) describing a particular class of effective holographic QFTs~\cite{McGough:2016lol,Hartman:2018tkw}. The latter arise as irrelevant deformations of CFT by an operator being a square of its energy-momentum tensor.

The plan of our paper is as follows: in section~\ref{sec:metrics} we will first review different forms for the metrics of AdS space and AdS black holes, which we will require in our later calculations. Section~\ref{sec:ads} is devoted to the calculation of the partition function with a three-sphere boundary. In section~\ref{sec:bhs} we will then extend these results by changing the boundary topology to $S^1 \times S^2,$ which will allow us to include black holes to our discussion. The connections with cosmology are discussed in section \ref{sec:cosmology} and an outlook and some interesting open problems are provided in section~\ref{sec:discussion}. 

\section{Useful metrics for AdS and black holes} \label{sec:metrics}

The evaluation of gravitational path integrals is greatly simplified by choosing particularly well adapted metric ans\"{a}tze, which differ from the metric forms that are most often used in other contexts. In this section, for convenience we will present the AdS and asymptotically AdS black hole metrics both in a common form and in the form that we will employ later. 

We will consider general relativity in four spacetime dimensions in the presence of a negative cosmological constant $\Lambda,$ with action
\begin{align}
S = \frac{1}{16\pi G} \int d^4x \sqrt{-g} \left[ R - 2 \Lambda \right]\,,
\end{align}
where one may define 
\begin{equation}
\Lambda \equiv - \frac{3}{l^2}
\end{equation}
with $l$ denoting the radius of curvature of the maximally symmetric AdS solution. We specialise here and in the following to gravity in four dimensions because we want to draw lessons about cosmology in our Universe. It would certainly be interesting to generalise the findings of our article to an arbitrary number of dimensions.

The Euclidean version of the empty AdS solution may be written as
\begin{align}
\label{eq.metricAdS}
ds^2 = d\rho^2 + l^2 \sinh^2 \left( \frac{\rho}{l}\right) d\Omega_3^2\,,
\end{align}
where $d\Omega_3^2$ is the metric on the unit three-sphere with volume~$V_3=2\pi^2$ and the (asymptotic) boundary resides at~$\rho \rightarrow \infty$. Via holography, the gravitational action evaluated on-shell on this solution and supplemented with appropriate counter terms to kill divergences incurred as $\rho \rightarrow \infty$ approximates the logarithm of a partition function (free energy) for a dual CFT living on the boundary. For the metric~\eqref{eq.metricAdS} the latter is a three-sphere. One reason why it is interesting to compute partition functions for three-dimensional QFTs on spheres stems from this quantity being a natural measure of the number of degrees of freedom in such QFTs~\cite{Jafferis:2011zi}.

We will find it useful to consider a metric of the form \cite{Halliwell:1988ik}
\begin{align}
\label{metric}
ds^2 = -\frac{N^2}{q(r)}dr^2 + q(r) d\Omega_3^2\,,
\end{align}
where $N$ denotes the lapse function. Note that the minus sign in the above equations is not a standard convention in holography, but it will facilitate the comparison with cosmology. Also, in the end we will define Eq.~\eqref{eq.Zgravity} as a path integral over complex geometries, so this is just a choice of a convention. Moreover we will consider situations in which there is a boundary at a fixed radius. For this purpose it will be useful to rewrite the solutions in terms of a radial coordinate with a finite range, say $0 \leq r \leq 1.$ A patch of the EAdS solution for $0 \leq \rho \leq \rho_{max} = l \, \mathrm{arcsinh}(\frac{R_{3}}{l})$ then corresponds to 
\begin{align}
N & = \pm i \, l\left( \sqrt{R_3^2+l^2} + l\right) \,, \\
q(r) &= \left(\sqrt{R_3^2 + l^2} + l \right)^2 r^2  - 2 l \left(\sqrt{R_3^2 +l^2} + l\right)r \,. \label{scalefactorads}
\end{align}
Note that the lapse function is imaginary, in accordance with the Euclidean nature of the solution. Here $R_3$ may be seen to fix the proper radius of the three-sphere at the outer boundary at coordinate location $r=1$. In particular, within this ansatz considering the solution all the way to the asymptotic boundary corresponds to blowing-up the radius $R_{3}$.

We will also consider metrics with $S^1 \times S^2$ topology on constant radial surfaces. A~corresponding metric for AdS space is
\begin{align}
ds^2 = \frac{d\rho^2}{\left( \frac{\rho^2}{l^2} + 1\right)} + \left( \frac{\rho^2}{l^2} + 1\right) d\tau^2 + \rho^2 d\Omega_2^2\,, \label{AdSmetric}
\end{align}
where 
\begin{equation}
d\Omega_2^2=d\theta^2 + \sin^2(\theta)d\phi^2
\end{equation}
is the metric on the unit two-sphere, and where we have chosen the ``time'' coordinate $\tau$ to be Euclidean. In these coordinates the empty Anti-de Sitter solution can be straightforwardly extended to include a (Euclidean) Schwarzschild black hole \cite{Carter:1973rla}, as
\begin{align}
ds^2 = \frac{d\rho^2}{\left( \frac{\rho^2}{l^2} + 1 - \frac{2M}{\rho}\right)} + \left( \frac{\rho^2}{l^2} + 1 - \frac{2M}{\rho}\right) d\tau^2 + \rho^2 d\Omega_2^2\,, \label{AdSbhmetric}
\end{align}
where $M$ denotes the mass of the black hole. The horizon radius $r_+$ of the black hole is given by the real root of 
\begin{align}
\frac{\rho^3}{l^2} + \rho -  2M=0\equiv \frac{1}{l^2}(\rho-r_+)(\rho-r_1)(\rho-r_2)\,, \label{cubic}
\end{align} 
while the other two roots $r_1,r_2$ form a complex conjugate pair, since the discriminant of this cubic equation is negative\footnote{The cubic roots have the properties that $r_++r_1+r_2=0$ and $r_1 r_2 = r_+^2 + l^2.$}. From this one immediately obtains an expression for the mass in terms of the horizon radius
\begin{align}
M=\frac{1}{2} r_+ \left( 1 + \frac{r_+^2}{l^2} \right)\,. \label{mass}
\end{align}
In order for the manifold to avoid a conical singularity at the horizon, one must impose that the $\tau$ coordinate is periodic (so that the near-horizon metric resembles that of the origin of flat space in polar coordinates) with period \cite{Hawking:1982dh}
\begin{align}
\beta = \frac{4\pi l^2 r_+}{3r_+^2+l^2}\,. \label{horizon}
\end{align}
The AdS/CFT correspondence maps the mass $M$ in~\eqref{mass} to the expectation value of the corresponding CFT Hamiltonian in a thermal state on a unit two-sphere at temperature equal to~$1/\beta$~\cite{Balasubramanian:1999re}.

Once again we would like to bring the metric \eqref{AdSbhmetric} into a form where the radial coordinate has finite range.  For this, we will first pick a radius $\rho=R_2,$ where $R_2$ denotes the radius of the two-sphere on the boundary. Our radial coordinate $r$ ranges from $0$ to $1,$ and should interpolate between $r_+$ and $R_2.$ Thus we will define
\begin{align}
\label{eq.defbBH}
\rho \equiv b(r) = r (R_2-r_+) + r_+.
\end{align}
Here one can see that sending $\rho$ to $\infty$ is equivalent to blowing up $R_{2}$, as we wrote earlier. Using the ansatz for the Kantowski-Sachs class of metrics~\cite{Halliwell:1990tu},
\begin{align}
\label{eq.bhmetric}
ds^2 = - \frac{b(r)}{c(r)} N^2  dr^2 + \frac{c(r)}{b(r)} d\tau^2 + b(r)^2 d\Omega_2^2\,,
\end{align}
the black hole geometry can now be rewritten as 
\begin{subequations}
\begin{align}
\label{eq.BHNc}
N & = \pm i (R_2 - r_+)\,, \\
c(r) & = \frac{1}{l^2}[b^3(r) + l^2 b(r) - r_+^3 - l^2 r_+]\,.
\end{align}
\end{subequations}
On the outer boundary at $r=1$ we have
\begin{align}
b(1) = R_2 \quad \mathrm{and} \quad c(1) = \frac{1}{l^2}(R_2^3 + l^2 R_2 - r_+^3 - l ^2 r_+ )\,. \label{b1c1}
\end{align}
If we denote the period of $\tau$ by $\beta$ then we can see that the size of the circle direction on the boundary is given by
\begin{align}
\sqrt{\frac{c(1)}{b(1)}}\, \beta \equiv R_1\,. \label{R1}
\end{align} 
Keeping $R_1$ and $R_2$ fixed specifies the size of the outer boundary. At the inner boundary at~$r=0,$ the metric \eqref{AdSbhmetric} implies
\begin{align}
b(0) = r_+\quad \mathrm{and} \quad c(0)=0\,.
\end{align}
The value $b(0)$ effectively specifies the mass of the black hole, according to \eqref{mass}. Again a conical singularity at $r=0$ is avoided provided the periodicity $\beta$ is given by
\begin{align}
\beta = \frac{4\pi b(0) |N|}{\dot{c}(0)} = \frac{4\pi l^2 r_+}{3 r_+^2 + l^2}\,, \label{noconical}
\end{align}
where a dot denotes a derivative w.r.t. $r.$ 

Pure Euclidean AdS space~\eqref{AdSmetric} is recovered in the limit $r_+ \to 0,$ and in that case the periodicity $\beta$ is arbitrary since the manifold is smooth in any case.


\section{$S^3$ boundary and Euclidean AdS$_4$ saddles} \label{sec:ads}

\subsection{Neumann condition at $r = 0$} \label{subsec:N}

We will first review how Euclidean AdS space is obtained as the saddle point of a gravitational path integral. This calculation was done previously by Caputa and Hirano in \cite{Caputa:2018asc}, and in three dimensions in \cite{Donnelly:2019pie}\footnote{See also Ref.~\cite{Hirano:2019szi} for another recent application of the minisuperspace approach in holography.}. Here we will perform the analogous calculation in a different style adopted from Ref.~\cite{Feldbrugge:2017kzv}, which has the advantage that it will allow us to extend the calculation to black holes in the next section. Also, motivated by the extension to black holes, we will impose here different boundary conditions at $r = 0$ than in \cite{Caputa:2018asc}. In subsection \ref{subsec:D} we will show how the results of~\cite{Caputa:2018asc} fit into our framework, and we will discuss some implications of our studies in \ref{subsec:lessons}.

The object we are interested in is the partition function
\begin{align}
\label{eq.defZ3}
Z(R_3) = \int^{R_3} d[g_{\mu\nu}] e^{\frac{i}{\hbar}S}\,,
\end{align} 
with a three sphere of radius $R_3$ at the fixed (outer) boundary. Note that the nature of the bulk metrics that we integrate over is going to be determined by the contour of integration in the lapse integral and will in general involve complex metrics. Also, the signature of the metric in which a dual QFT lives is fixed by the outer boundary condition and unaltered by this genuinely bulk phenomenon. To be more precise about our aim, we want to define Eq.~\eqref{eq.defZ3} within the minisuperspace approach so that it is mathematically meaningful and has features consistent with calculating a partition function in a dual QFT.

The action we will consider consists of the Einstein-Hilbert action with a negative cosmological constant, supplemented by the Gibbons-Hawking-York (GHY) surface term~\cite{York:1972sj,Gibbons:1976ue} at the (outer) boundary,
\begin{align}
\label{eq.SgravNeumann}
S = \frac{1}{16\pi G}\int d^4 x \sqrt{-g} \left[ R + \frac{6}{l^2} \right] + \frac{1}{8\pi G}\int_{outer} d^3 y \sqrt{h} K +S_{ct} \,,
\end{align}
where the counterterms $S_{ct}$ will be discussed below. Note that in this subsection we do not add any surface terms on the inner boundary, for reasons that will become clear. We work with minisuperspace metrics given the ansatz~\eqref{metric}. The coordinate $r$ interpolates between the inner boundary at $r=0$ and the outer boundary at $r=1,$ that is to say $0 \leq r \leq 1.$ Here $N(r)$ is the lapse function and $q(r)$ is the scale factor squared, which determines the size of the three-sphere. We will denote $q(r=0) \equiv q_0$ and $q(r=1)\equiv q_1=R_3^2.$ The reason for choosing this less familiar metric ansatz is that the action ends up being quadratic in $q,$ which will be very useful in evaluating the path integral over $q.$ In fact the action reduces to\footnote{One subtlety that we want to highlight is that in passing from the general action~\eqref{eq.SgravNeumann} to its form for the minisuperspace metrics~\eqref{metric} we made a choice of a branch in the expression~$\sqrt{-g}$. For purely Euclidean metric, this would be the standard choice one makes.} 
\begin{equation}
S = \frac{3\pi}{4 G} \int_0^1 dr \left[ - \frac{\dot{q}^2}{4 N} + N \left(1 + \frac{q}{l^2}\right) \right] - \frac{3 \pi q_0 \dot{q_0}}{8 GN} + S_{ct}\,, \label{NDaction}
\end{equation}
where a dot over a function denotes here and in the following a derivative w.r.t. $r$. Second derivatives acting on $q$ have been eliminated using integration by parts, and the resulting surface term at $r=1$ has eliminated the GHY surface term there while introducing a surface term~$- \frac{3 \pi q_0 \dot{q_0}}{8 GN}$ at $r=0.$ Variation of the action w.r.t. $q$ leads to 
\begin{align}
\delta S = \frac{3\pi}{4G} \int_0^1 dr \frac{\delta q}{2N} \left[  \ddot{q} + \frac{2 N^2}{l^2} \right] - \frac{3 \pi\dot{q}_1 \delta\left(q_1\right)}{8GN} - \frac{3 \pi q_0 \delta\left(\dot{q}_0\right)}{8GN} + \delta S_{ct}\,. 
\end{align}
Thus we obtain the equation of motion $ \ddot{q}  = - \frac{2 N^2}{l^2}\,, $ supplemented by the boundary conditions that we can hold $q$ fixed at the outer boundary, as desired (the variation $\delta S_{ct}$ will be consistent with this), while self-consistency upon not including the Gibbons-Hawking term on the inner boundary of the integration range forces us to fix $\dot{q}$ there. More properly we should say that it is the momentum conjugate to the scale factor, 
\begin{equation}
\Pi = \frac{\delta \mathcal{L}}{\delta \dot{q}} = -  \frac{3\pi}{8GN}\dot{q}\,,
\end{equation}
that will be fixed on the inner boundary. This Neumann condition, understood as fixing the momentum to some value (not necessarily 0), is not strictly needed in the present calculation: as we will show in the next subsection, we could have used the Dirichlet condition here. However, a momentum condition will be necessary when including black holes, if we want to reproduce standard results of black hole thermodynamics in the canonical ensemble and we interpret the result of the bulk path integral as the thermodynamic partition function. It will be useful for us to parameterise the momentum at $r=0$ by a re-scaled parameter $\alpha,$ 
\begin{equation}
\Pi_0  = -  \frac{3\pi}{8GN}\dot{q}_0 \equiv -  \frac{3\pi}{4G}\alpha\,. \label{momentumcondition}
\end{equation}
To proceed, we must evaluate the path integral over the metric i.e. the integrals over the lapse and the scale factor, $\int d[g_{\mu\nu}] = \int d[N]d[q],$ where we will ignore Jacobian factors given that we will eventually evaluate the partition function in the saddle point approximation. Here and in the following we will make use of an old result of~\cite{Teitelboim:1981ua,Halliwell:1988wc}, namely that one can use the gauge freedom of general relativity to restrict the sum to run over manifolds in which the lapse $N$ does not depend on~$r$. This drastically simplifies the integral over $q$ and also transforms the functional integral over $N$ into an ordinary integral.  The procedure to evaluate the path integral over $q$ is to shift variables by writing $q = \bar{q} + Q$ \cite{Halliwell:1988ik}. Here $\bar{q}$ denotes a solution of the equation of motion for $q$ respecting the boundary conditions $\Pi(0)=\Pi_0$ and $q(1)=q_1=R_3^2.$ Explicitly, we have \begin{align}
\bar{q}(r) = - \frac{N^2}{l^2} (r^2-1)  +2N\alpha (r-1)  + R_3^2\,. \label{saddlescale}
\end{align}
Meanwhile $Q$ is an arbitrary perturbation (not necessarily small), which obeys the boundary conditions, i.e. vanishes at $r = 1$ and has a vanishing momentum at $r = 0$. Since the action is quadratic in $q,$ the path integral then turns into a factor given by the action evaluated along the $\bar{q}$ plus a Gaussian integral over $Q$. As we show in appendix \ref{sec:determinant}, this yields just a numerical prefactor. Thus, up to this numerical factor, we are left with
\begin{subequations}
\begin{align}
Z(R_3) & = \int dN e^{i(S_0+S_{ct})/\hbar}\,, \label{lapsepartition}\\
\frac{4G}{\pi} S_0(N)&=  \frac{N^3}{l^4}-3\alpha \frac{N^2}{l^2} +3N\left(\frac{R_3^2}{l^2}+1+\alpha^2 \right) -3\alpha R_3^2 \,, \label{sphereaction}
\end{align}
\end{subequations}
where it is important that the counter terms contain no dependence on the lapse. 
While the integral over $N$ from $- i\infty$ to $0$ converges in this particular case, below we take a route of Ref.~\cite{Feldbrugge:2017kzv} which will apply to all the cases considered in the present work\footnote{We will see below that $N=0$ is not a natural end point of integration, as the relevant steepest descent contour continues beyond this point.}.
 
\begin{figure}[h]
\includegraphics[width=0.5\textwidth]{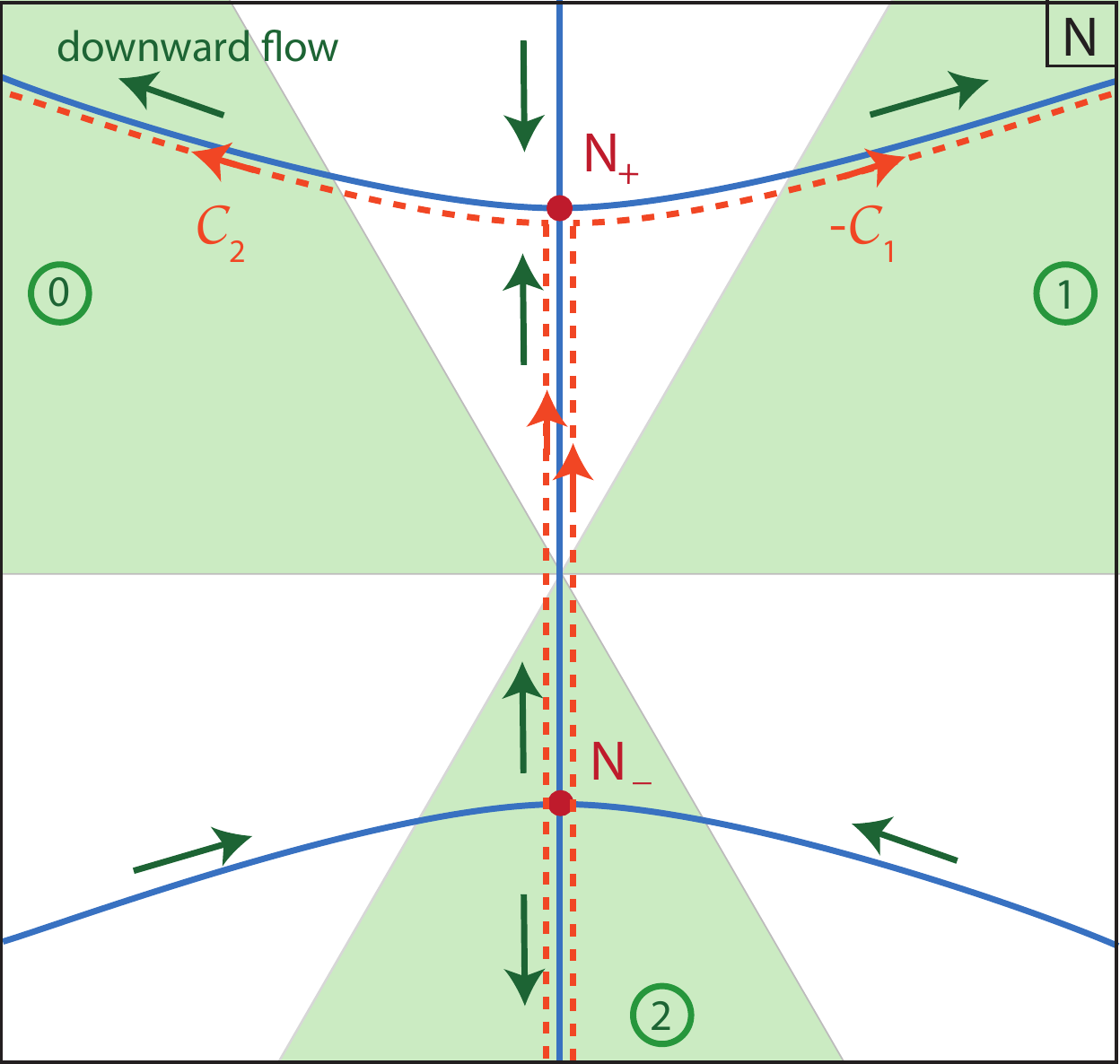}
\caption{The figure shows the structure of the flow lines with boundary conditions $\alpha=+i$, $q_1 = R_3^2$.  The saddle point $N_-$ represents the EAdS geometry, while $N_+$ represents a singular section of complexified AdS space. The asymptotic regions of convergence are shown in light green, and are labelled by the encircled numbers $0,1,2.$ We denote the contour of integration linking region $0$ to region $1$ by ${\mathcal C}_0 = 0 \to 1,$ and similarly ${\mathcal C}_1 = 1 \to 2$ and ${\mathcal C}_2 = 2 \to 0.$ Integration along ${\mathcal C}_0$ yields the Airy Ai function. Meanwhile, the combination ${\mathcal C}_2 - {\mathcal C}_1$ gives a result proportional to the Airy Bi function. This sum of contours is equivalent to summing the two contours shown by the orange dashed lines, which run via the saddle point $N_-$ to the saddle point $N_+$ and from there to opposite regions of convergence. Note that the required contours of integration are neither Lorentzian nor Euclidean and the combination ${\mathcal C}_2 - {\mathcal C}_1$ is the closest it gets to an effectively Euclidean path integral.} \label{fig:flowAdS}
\end{figure}

The lapse integral above can in fact be evaluated exactly. It is easiest to see this by shifting the integration variable to $\tilde{N} \equiv N-\alpha l^2,$ which leads to
\begin{align}
\label{eq.shiftedaction}
    \frac{4 G}{\pi} S = \frac{1}{l^4}\left[\tilde{N}^3 + 3 l^2 \left(R_3^2 + l^2 \right)\tilde{N} + \alpha (3+\alpha^2)l^6 \right]\,.
\end{align}
Then our path integral can be identified as an Airy integral (using $d\tilde{N}=dN$). The possible integration contours are defined in terms of the asymptotic regions of convergence labelled $0,1$ and $2$ in Fig. \ref{fig:flowAdS}. They are located at phase angles $\theta \equiv \arg(N)$: $0 \leq \theta \leq \frac{\pi}{3}$ (region $1$), $ \frac{2\pi}{3}\leq \theta \leq \pi$ (region $0$) and $ \frac{4\pi}{3}\leq \theta \leq \frac{5\pi}{3}$ (region $2$). We define the contours as ${\mathcal C}_0 = 0 \to 1$, ${\mathcal C}_1 = 1 \to 2$ and ${\mathcal C}_2 = 2 \to 0$, which define for us the two Airy functions as follows~\cite{Vallee},
\begin{equation}
\label{eq.defAiry}
Ai[z]=\frac{1}{2 \pi} \int_{\mathcal{C}_0}   dx \,  e^{ i \left(\frac{x^3}{3} + z x \right)} \quad \mathrm{and} \quad Bi[z]=i \frac{1}{2 \pi} \int_{\mathcal{C}_2 - \mathcal{C}_1 }   dx \,  e^{ i \left(\frac{x^3}{3} + z x \right)}\,.
\end{equation}

Since the CFT partition function is real, we should also expect the gravitational partition function to yield a real valued result. There are then two possible contours. The first is to integrate along ${\mathcal C}_0,$ and this yields a result proportional to the Airy function $Ai\left[ \left(\frac{3\pi l^2}{4 G \hbar} \right)^{\frac{2}{3}} \left( R_3^2+l^2\right)\right].$ At large $R_3$ this function is exponentially suppressed $\sim e^{-R_3^3}$ and thus cannot represent the desired answer. The other possibility is to integrate along ${\mathcal C}_2 - {\mathcal C}_1.$ This can be seen as follows: our integrand is odd in the lapse, which implies that under the transformation $N\to - N^*$ the integrand changes into its complex conjugate. An integration contour that is even under $N \to -N^*$ can then be split up into two pieces which are reflection symmetric across the imaginary $N$ axis, and, if they have the same orientation (e.g. left to right), they represent a sum of an integral and its complex conjugate -- thus, the result is real. This is the case for the contour ${\mathcal C}_0 = -({\mathcal C}_2 + {\mathcal C}_1),$ which gives the Airy Ai function. The other possible contour that is even under $N \to - N^*$ is the difference ${\mathcal C}_2 - {\mathcal C}_1.$ It corresponds to the difference of an integral and its complex conjugate and is pure imaginary; after multiplication with the imaginary unit $i$, it also yields a real result, namely the Airy Bi function.

Reinstating the spatial volume $V_3$ and using Eq.~\eqref{eq.defAiry}, the integral over ${\mathcal C}_2 - {\mathcal C}_1$ yields the following answer  
\begin{align}
    Z(R_3) &= e^{i \frac{V_3}{8\pi G \hbar}\alpha (3+\alpha^2)l^2} \, Bi\left[ \left(\frac{3V_3}{8\pi G \hbar l} \right)^{\frac{2}{3}} \left( R_3^2+l^2\right)\right] e^{\frac{i}{\hbar}S_{ct}}\,. \label{Z3result}
\end{align}
We will determine $\alpha$ momentarily, but first it is useful to look at the large $R_3$ limit of this expression. Naively our result diverges as the boundary is pushed to infinity, $R_3 \rightarrow \infty$. This is the usual infinite volume divergence found in the context of asymptotically AdS spacetimes. This divergence is cured by the introduction of counter terms~\cite{Balasubramanian:1999re,deHaro:2000vlm},
\begin{equation}
S_{ct} = \frac{i}{16\pi G} \int_{outer} d^3 y \sqrt{h} \left(\frac{4}{l} + l \,  R^{(3)}\right),
\end{equation}
where $h$ and $R^{(3)}$ are the determinant and the Ricci scalar of the three-metric on the outer boundary. For the metric ansatz \eqref{metric} they become $S_{ct} = +\frac{iV_3}{8 \pi G l} (2 R_3^3 + 3 R_3 l^2)\,,$ so that\footnote{Note that the counter terms depend only on the radius $R_3$ and not on its derivative, and hence the variation of the counter terms is consistent with our Dirichlet condition at $r=1.$} 
\begin{align}
e^{\frac{i}{\hbar}S_{ct}} = e^{-\frac{V_3}{8\pi G \hbar l} (2 R_3^3 + 3 R_3 l^2)}\,. \label{ct}
\end{align}
For large values of $R_{3}$ the gravitational path integral~\eqref{Z3result} takes the form
\begin{align}
    Z(R_3)  \approx e^{\frac{V_3}{8\pi G \hbar} \left[2 \left( R_3^2 + l^2\right)^{3/2} + i \alpha (3+\alpha^2)l^2 -2R_3^3 - 3 l^2 R_3\right]},
\end{align}
where we kept terms up to ${\cal O}(1/R_{3})$ in the exponent, and leads to
\begin{align}
\label{eq.ZS3Neumannexact}
Z = e^{i \frac{V_3}{8\pi G \hbar}\alpha (3+\alpha^2)l^2}
\end{align}
as $R_3 \to \infty$. We want to emphasise that, up to an ambiguity in the notion of the path integral measure, this is the exact result of the path integral within the minisuperspace ansatz.

On the QFT side of holography, the partition function on a three-sphere can be evaluated \emph{exactly} in the very special case of the ABJM theory~\cite{Aharony:2008ug} using localisation~\cite{Pestun:2007rz}. The result reads~\cite{Fuji:2011km,Marino:2011eh}
\begin{align}
\label{eq.ZABJM}
Z = Ai\left[ \left(\frac{3V_3l^2}{8\pi G \hbar}\right)^{2/3}\right],
\end{align}
where we utilised the holographic dictionary for the ABJM theory to reinstate $G$ instead of the number of underlying QFT degrees of freedom. In the limit when the gravity side is described in terms of classical gravity, i.e. when the argument in eq.~\eqref{eq.ZABJM} is very large, one gets
\begin{align}
Z \sim e^{-2V_3l^2/(8\pi G \hbar)}.
\end{align}
Thus we see that we recover the leading term of the ABJM result with the choice $\alpha=+i.$ At the current point in the calculation, this choice appears somewhat mysterious, but we will see shortly that it has a perfectly sensible physical origin. There are two immediate consequences however: the first is that this identification means that the momentum condition $\alpha=+i,$ corresponding to $\Pi_0 = - \frac{3\pi}{4G} i,$ must be fixed and should not be summed over in the partition function. The second is that with $\alpha=+i,$ the total partition function in Eq. \eqref{Z3result} is real for any value of $R_3,$ as expected on general grounds.

\begin{figure}[h]
\includegraphics[width=0.5\textwidth]{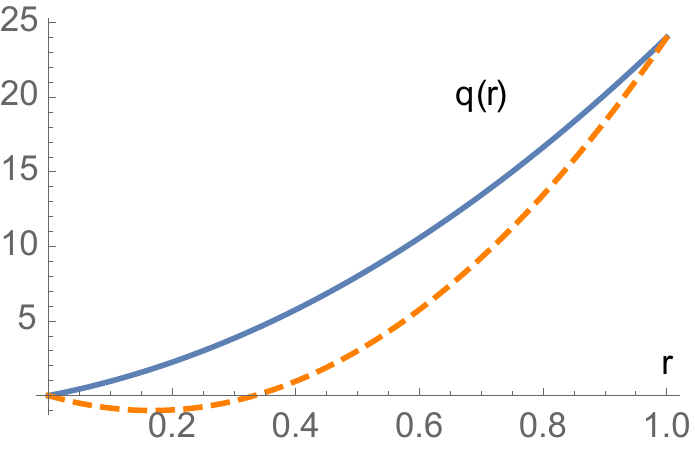}
\caption{This graph shows the profile of the scale factor squared $q(r)$ at the saddle points $N_- = -4i, N_+= 6i,$ obtained with the parameter values $l=1, R_3^2=24.$ The solid curve is given by $q(r)=8r+16r^2$ and represents EAdS space at $N_-$, while $N_+$ corresponds to a section $q(r)=12r+36r^2$ of (complexified) AdS space in which the scale factor turns imaginary in one region.} \label{fig:saddles1}
\end{figure} 

To gain more insight into this calculation, we will also perform the lapse integral \eqref{lapsepartition} in the saddle point approximation. For this it is useful to first study the nature of the saddle points. These are located at extrema of $S_0(N),$ i.e. at
\begin{subequations}
\begin{align} 
    N_\pm & = \alpha l^2 \pm i l \sqrt{R_3^2 + l^2}\,, \label{S3saddles} \\
    \frac{4 G}{\pi} S_0(N_\pm) &= \alpha (3+\alpha^2) l^2 \pm \frac{2i}{l}\left(R_3^2 + l^2 \right)^{3/2}\,, \label{saddleactionN}
\end{align}
\end{subequations}
where we also indicated the value of the action at the saddle points. We can obtain the saddle point geometry by inserting $N_\pm$ into Eq. \eqref{saddlescale},
\begin{align} \label{saddlemetric}
    \bar{q}(r)\mid_{N_\pm} = - \left( \alpha l \pm i \sqrt{R_3^2 + l^2}\right)^2 r^2 + 2\alpha \left( \alpha l^2 \pm i l \sqrt{R_3^2 + l^2}\right) r - l^2 (1+\alpha^2)\,.\end{align}
Here we can see that if we are to evade a physical boundary at $r=0,$ we need to restrict to the momentum conditions $\alpha=\pm i,$ so that $\bar{q}(0)=0.$ This consideration already reduces the possible values of $\alpha$ to just two. All saddle points consist of sections of complexified AdS spacetime. When $\alpha=+i,$ the saddle point $N_-$ corresponds to the usual Euclidean AdS space, which we expected to find. The other saddle point, $N_+,$ describes a section of Euclidean AdS glued onto a reversed-signature EAdS piece. For this last saddle point, the squared scale factor~$\bar{q}$ passes through zero and becomes imaginary, and thus we would expect perturbations to blow up there, cf. Fig. \ref{fig:saddles1}. Also note from Eq. \eqref{saddleactionN} that the EAdS solution has a higher weighting than the singular saddle point. By contrast, when $\alpha =-i$ the EAdS is at $N_+$ and the singular geometry at $N_-,$ and in this case the singular geometry dominates. In the limit of large $R_3$ the subdominant saddle points are suppressed, which indeed implies that we should choose $\alpha=+i.$ 

The saddle points, along with their steepest descent lines, are shown in Fig. \ref{fig:flowAdS}. The contour of integration ${\mathcal C}_2 - {\mathcal C}_1,$ which we chose above, can then be deformed into the sum of two contours that are symmetric w.r.t. the imaginary lapse axis, and which run from negative imaginary infinity either to the convergence region $0$ or $1.$ These contours follow the steepest descent path through the saddle point $N_-$ representing EAdS space, on to the saddle point $N_+$ and from there along either half of the steepest descent path associated with $N_+.$ At $N_+$ the two parts of the total integration contour run parallel to the real lapse axis, but in opposite directions, implying that the end result will not contain a contribution from the singular saddle point. In the saddle point approximation, including the counter terms, the partition function is then approximated as
\begin{align}
Z(R_3) & \approx e^{\frac{V_3}{8\pi G \hbar l}\left(-2l^3 + \frac{3l^4}{4R_3} + {\cal O}(R_3^{-3})\right)} \,,
\end{align}
in agreement with our earlier result~\eqref{eq.ZS3Neumannexact} for $\alpha = +i$.

We are thus able to define a partition function peaked around Euclidean AdS space, by using a Neumann condition at $r = 0$ and a Dirichlet condition at the boundary. Perhaps the most surprising aspect of this calculation is that the contour for the lapse integral can be neither Euclidean nor Lorentzian, but must be inherently complex, as shown in Fig. \ref{fig:flowAdS}. However, it is interesting to note that if one were to ``sum'' the contours together, then ${\mathcal C}_2 - {\mathcal C}_1$ is in fact equal to the Euclidean lapse axis\footnote{Note that the result of this calculation differs from the naive summation over the negative imaginary axis that one can do in this special case.}. This may be the closest one is able to come to a realisation of Euclidean quantum gravity with the caveat that we discussed before.

We will see later that many of these aspects persist when we extend our calculation to include black holes. For now, we will first compare our calculation with one using Dirichlet boundary conditions on both ends.


\subsection{Dirichlet condition at $r = 0$} \label{subsec:D}

To compare with Ref.~\cite{Caputa:2018asc} we have to compare our results with the calculation performed with Dirichlet boundary conditions
\begin{align}
q(r=0) = 0\,, \qquad q(r=1)=R_3^2\,.
\end{align}
Note that in this case the condition of starting at zero size is put in from the outset\footnote{As we have already mentioned, starting with a non-zero size would superficially imply including another holographic QFT. This is inconsistent with consideration of a partition function, hence our prescription.}. It will thus hold everywhere, i.e. also off-shell, and not just at the saddle points. However, this condition does not guarantee that at $r=0$ the geometry will be regular -- in fact it will only be so at the saddle points. For the Dirichlet calculation, we must use the Einstein-Hilbert action supplemented with the GHY terms at both $r=0,1,$ which reduces to the minisuperspace action\footnote{The Dirichlet condition $q_0=0$ is special in the sense that the surface term vanishes for this particular value, cf. Eq.~\eqref{NDaction}, hence one does not necessarily need the GHY term at $r=0.$ However, if one thinks of this calculation as integrating from smaller and smaller initial sizes, then it makes sense to add the GHY term in order to ensure a smooth limit when $q_0 \to 0.$}
\begin{equation}
\frac{8\pi G}{V_3}S = 3  \int_0^1 dr \left[ - \frac{\dot{q}^2}{4 N} + N (1 + \frac{q}{l^2}) \right]\,.
\end{equation}
The second term in the action arises from the positive curvature of the three-sphere, and the last term from the cosmological constant. The GHY boundary term has eliminated all second derivatives, so that the variational problem will be well posed when imposing Dirichlet boundary conditions on $q.$ The trick to evaluate the path integral over $q$ is once again to shift variables by writing $q = \bar{q} + Q.$ Here $\bar{q}$ denotes a solution of the equation of motion for $q$ respecting the boundary conditions,
\begin{align}
\bar{q}(r)= - \frac{N^2}{l^2} r^2 + \left(\frac{N^2}{l^2} + R_3^2\right) r\,.
\end{align}
Meanwhile $Q$ is an arbitrary perturbation (not necessarily small) with vanishing value at the end points $Q(0)=Q(1)=0.$ Since the action is quadratic in $q,$ the path integral then turns into an integral over $\bar{q}$ which is just a given function of $r$ and can be integrated directly, plus a Gaussian integral over $Q$ which just changes the prefactor by a factor $1/\sqrt{N}$~\cite{Halliwell:1988ik}. Thus, up to an overall numerical factor that we were persistently neglecting throughout the text, we are left with
\begin{subequations}
\begin{align}
\Psi & = \frac{e^{i \frac{\pi}{4}}}{\sqrt{\pi \hbar}} \int \frac{dN}{\sqrt{N}} e^{iS_0/\hbar}\,, \label{lapseintegralDirichlet} \\
\frac{8\pi G}{3V_3} S_0 &=  \frac{N^3}{12 l^4} + \frac{N}{2 l^2}(R_3^2 +  2 l^2)  - \frac{R_3^4}{4 N} \,.
\end{align}
\end{subequations}
Here we have denoted the path integral by the new letter $\Psi,$ since the relation to the partition function of the previous section is a priori not clear. The asymptotic convergence regions at infinity are unchanged from the Neumann case, but in addition the action now contains a pole at $N=0,$ so that there are additional choices for the lapse integration contour. Intuitively the appearance of a singularity at $N = 0$ should not be all that surprising. In fact here we are summing over four-geometries which interpolate between two three-spheres of radii $q(r=0)=0$ and $q(r=1)=R_3^2>0$. When $N = 0$, then the proper distance between them vanishes and the singularity is signaling that the corresponding geometry is not smooth. Note also that this pole invalidates any attempts to perform the integral along the negative imaginary axis within the Euclidean quantum gravity approach. This singularity was not present in the case where we fixed the momentum at $r=0$ as this Neumann condition can be thought of as a sum over all possible sizes $q_0$. Indeed from Eq. (\ref{saddlescale}) one can see that the size of the sphere located at $r=0$ changes with the lapse and the geometry with $N=0$ is regular and has $q(r=0)=q(r=1)$.

\begin{figure}[h]
\includegraphics[width=0.8\textwidth]{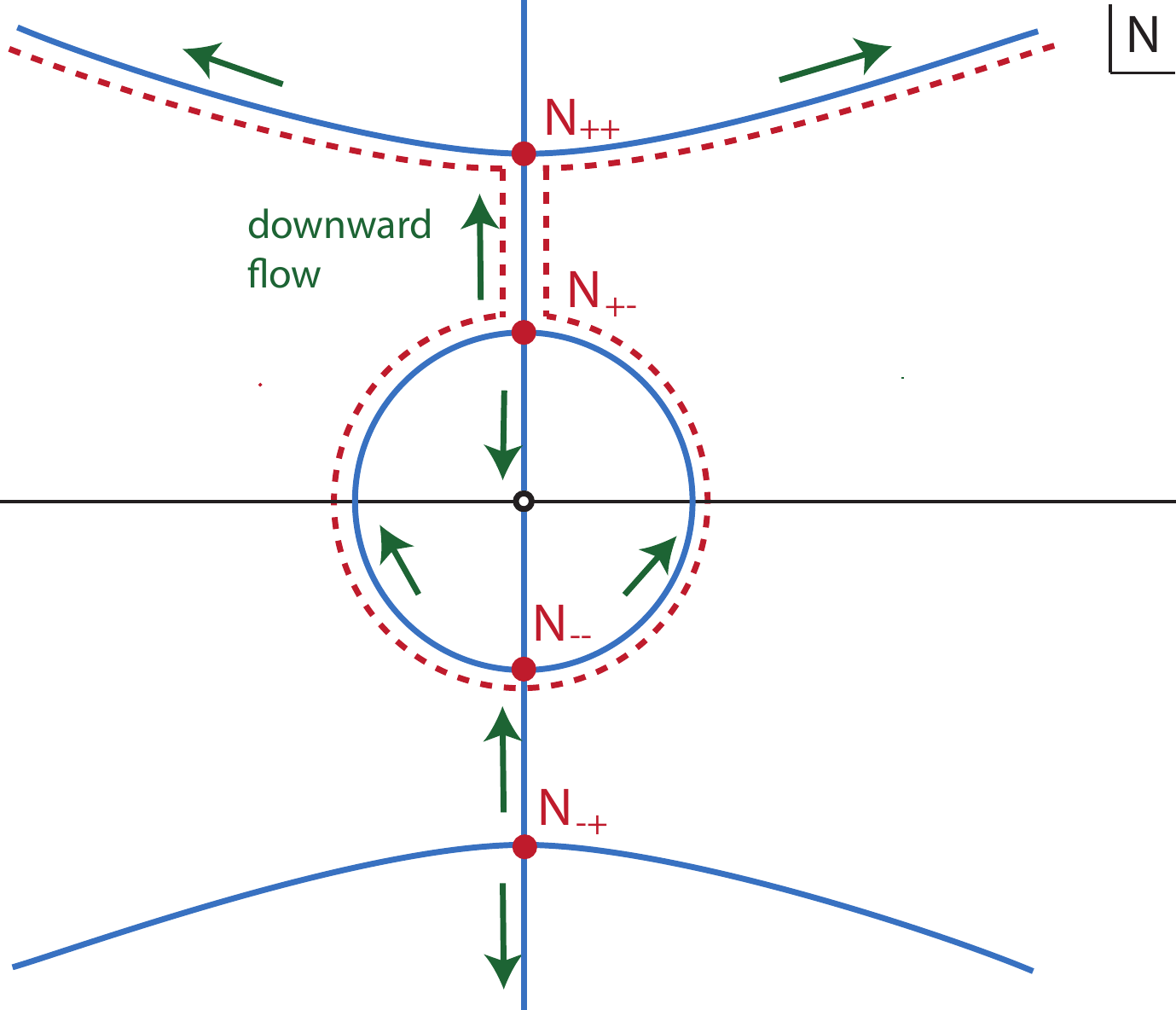}
\caption{Flow lines in the complexified plane of the lapse function, for Dirichlet boundary conditions. The saddle points closer to the origin have $c_2=-1,$ while those two that are further away have $c_2=+1.$ The dashed line in the figure indicates the preferred contour of integration. Note that the action contains a singularity at $N=0.$} \label{fig:flows}
\end{figure}

It turns out that also in the present case the lapse integral in Eq.~\eqref{lapseintegralDirichlet} can be evaluated exactly~\cite{Vallee}. The trick is to rewrite the measure factor as a Gaussian integral $e^{-i \,\frac{(q_1 - q_0)^2}{4 N}} e^{i\,\pi/4}\sqrt{\frac{\pi}{N}} = \int d\xi \, e^{i \,N\,\xi^2 +i \,(q_1 - q_0) \,\xi},$ and then, after a change of variables to $N \pm 2 \xi$, the integral \eqref{lapseintegralDirichlet} can be identified as a product of two Airy integrals. Thus the solution is given by the product of two Airy functions. The choice of integration contour for the lapse determines the type of Airy functions, where care is needed to ensure that all integrals converge. We once again require the resulting quantity to be real in order to interpret it as a partition function, and moreover, this time it must be symmetric w.r.t. the inner and outer boundaries since we imposed Dirichlet conditions on both ends. One possibility is to consider the real contour for the lapse running above the origin; correspondingly $\xi$ runs along the real axis and the path integral is given by $\Psi \propto Ai\left[\left(\frac{3V_3l^2}{8\pi G \hbar}\right)^{2/3}\right]Ai\left[\left(\frac{3V_3}{8\pi G \hbar l}\right)^{2/3}\left(R_3^2 + l^2\right)\right]e^{\frac{i}{\hbar}S_{ct}} $. This choice would however give a vanishing result in the limit where the boundary is pushed to infinity $R_3 \rightarrow \infty$. The solution with the right asymptotic behaviour is then given by
\begin{align}
\Psi \propto & \,\,\, Ai\left[\left(\frac{3V_3l^2}{8\pi G \hbar}\right)^{2/3}\right]Bi\left[\left(\frac{3V_3}{8\pi G \hbar l}\right)^{2/3}\left(R_3^2 + l^2\right)\right]e^{\frac{i}{\hbar}S_{ct}} \nonumber \\ & + Bi\left[\left(\frac{3V_3 l^2}{8\pi G \hbar }\right)^{2/3}\right]Ai\left[\left(\frac{3V_3}{8\pi G \hbar l}\right)^{2/3}\left(R_3^2 + l^2\right)\right] e^{\frac{i}{\hbar}S_{ct}}, \label{Dresult}
\end{align}
where we have also added the counter terms. Obtaining this solution requires an integration contour for the lapse which runs along the real $N$ line, but passes below the singularity at $N=0,$ cf. Fig.~\ref{fig:flows}. This will become very clear when considering the saddle point approximation below. 

In the result above, the second line is suppressed compared to the first. In fact in the infinite $R_3$ limit, the second line disappears completely, and in the first line the counter term compensates for the $Bi$ function, leaving the result
\begin{align}
    \Psi \to Ai\left[ \left(\frac{3V_3 l^2}{8\pi G \hbar}\right)^{2/3}\right]\qquad \left(\text{as }\,\,\, R_3 \to \infty \right).
\end{align}
Thus the Dirichlet calculation reproduces the exact ABJM result obtained in a superconformal field theory -- see the discussion around Eq.~\eqref{eq.ZABJM}. This is rather surprising and is likely a coincidence, as our AdS calculation included only pure gravity, and was restricted to minisuperspace metrics. By comparison, the Neumann calculation in the same setting reproduced only the leading semi-classical term, which is truly all one could have hoped to recover in any case. As we will see below, the interpretation of the Dirichlet calculation is not entirely straightforward, as it does not reproduce the canonical ensemble once black holes are included. Still, it is interesting that it reproduces the associated CFT sphere partition function so precisely. This was already noticed in the work by Caputa and Hirano~\cite{Caputa:2018asc}.

As in the Neumann case, we can gain a little more insight into the Dirichlet calculation by evaluating the lapse integral in the saddle point approximation. There are now four saddle points, residing at the values 
\begin{equation}
N_{c_1 , c_2} = i \,  l \, c_1  [\sqrt{R_3^2 + l^2} + c_2 l]\,, \qquad c_1,c_2 =\pm 1\,.
\end{equation}
Thus all four saddle points reside on the imaginary axis. Note that we started with a Lorentzian metric ansatz in \eqref{metric}, but the lapse function at the saddle points nevertheless ends up being imaginary. Thus the saddle point geometries are Euclidean. In fact these are the four saddle points that we obtained in the calculation with Neumann conditions $\alpha=\pm i.$ Here they all appear together, because all four saddle points respect $q(0)=0.$ The two saddle points with $c_2 = 1$ are the singular bouncing solutions with the scale factor passing through zero. The two with $c_2 = -1$ are the EAdS geometries. The action at the saddle points once again reads
\begin{equation}
\frac{8\pi G}{V_3}S(N_{c_1 \, c_2}) = + c_1 \frac{2 i }{l} [(R_3^2 + l^2)^{3/2} + c_2 l^3 ]\,. \label{saddleactionD}
\end{equation}
It is purely imaginary. The saddle points with $c_1=+1$ will correspond to a suppressed weighting, while those with $c_1=-1$ will have an enhanced weighting, compared to a classical solution which would have a real action. 

To see which saddle points are relevant to the path integral, we must analyse the upwards/downwards flow lines, i.e. the steepest ascent/descent lines of the weighting, and moreover we still have to specify the contour of integration for the lapse function. The flow lines are shown in Fig. \ref{fig:flows}. Even though we have obtained the same saddle points as in the Neumann calculation, the flow lines are different, not least because there is now a singularity of the action at $N=0,$ which acts as an essential singularity from the point of view of path integration.

All four saddle points are linked by steepest ascent/descent lines. There are several options for the contour of integration, though, as we have already stressed, one cannot define a Euclidean path integral over the negative imaginary axis, which would diverge due to the singularity at $N = 0$. One can however define integrals along the Lorenztian line of real $N$ values, but one must choose whether to pass above or below the singularity at the origin. Another option is to consider contours that run from the region of convergence at negative imaginary infinity out to the region of convergence between $0$ and $\pi/3$ radians, or between $2\pi/3$ and $\pi$. Then both saddle points in the lower half plane would be relevant. Let us try to figure out which contour is the most sensible by comparing again to the expected QFT result~\eqref{eq.ZABJM} in the semi-classical limit,  which reduces to $e^{-\frac{2V_3l^2}{8\pi G \hbar}}\,.$ Now recall that from Eq.~\eqref{saddleactionD}, the saddle point approximation to the path integral will be given by a sum over terms of the form
\begin{align}
e^{-\frac{2V_3 c_1}{8\pi G \hbar }\left[(R_3^2+l^2)^{3/2} +c_2 l^3\right]} \approx e^{-\frac{V_3 c_1}{l}\left[2 c_2 l^3 + 2 R_3^3 + 3 R_3 l^2 + {\cal O}(\frac{1}{R_3})\right]}
\end{align}
where we have set $q_0=0.$ The counterterm is $e^{\frac{i}{\hbar}S_{ct}} = e^{-\frac{V_3}{8\pi G \hbar l} (2 R_3^3 + 3 R_3 l^2)}\,.$ The divergence in the two saddle points in the lower half plane is then cancelled by the counterterm. To match the expected Airy function result (providing us with the correct leading order results), we must also have $c_2=-1.$ Thus we have to pick the third saddle from the top, i.e. the upper one in the lower half plane, which we called $N_{--}$. Reproducing this result requires using the contour over complexified metrics passing below the origin. In Fig.~\ref{fig:flows} we denoted a sample contour with a red dashed curve. Note that the contours that originate at negative imaginary infinity cannot be used, as they would also pick up the singular saddle point $N_{-+},$ which moreover would lead to a mismatch with the expected QFT result.

\subsection{Comments} \label{subsec:lessons}

Before we move on to considering black hole spacetimes in the next section, let us pause here and summarise the most salient features encountered so far in our exploration. To start with, if one were to trust the minisuperspace path integral as an exact statement, then one is either naturally (the Neumann case in section~\ref{subsec:N}) or necessarily (the Dirichlet case in section~\ref{subsec:D}) led to integrate over complex metrics, also in the case of AdS quantum gravity. Such calculations require then an additional input regarding which contours in the space of complexified metrics to choose. These choices lead to different semi-classical limits and only some reproduce dual QFT expectations, such as the reality of the Euclidean path integral in dual QFT situations of interest or exact QFT results dictated by symmetries. It would clearly be desirable to have an entirely gravitational consistency criterion for a definition of the gravitational path integral, but for the moment we do not have one. 

Regarding more detailed findings in the two cases we consider, the results eventually gave rise to the same relevant saddle point upon adjusting the integration contours appropriately, but otherwise the two calculations differ. For example, the path integration measure is different in both cases, cf. Eq.~\eqref{lapsepartition} vs.~Eq.~\eqref{lapseintegralDirichlet}. While the Dirichlet calculation can be made to match the exact ABJM result~\cite{Fuji:2011km,Marino:2011eh}, see Eq.~\eqref{eq.ZABJM}, as noted earlier in Ref.~\cite{Caputa:2018asc}, we believe this agreement is accidental. In particular, ambiguities in the integration measure, stemming from our uncertainty about the fundamental definition of an integration measure over metrics (which are the analogue of ordering ambiguities in the associated Wheeler-DeWitt equations), alter the answer beyond the leading semi-classical exponent. Still, there exists a rather close link between the Neumann exponential \eqref{eq.ZS3Neumannexact} and the Dirichlet Airy function \eqref{Dresult}, which stems from the fact that the Fourier transform of the Airy function is indeed an exponential with a cubic exponent. More precisely, we have that \cite{Vallee}
\begin{align}
    \int_{- \infty}^{\infty} dq_0\, e^{i \, q_0 \, \Pi_0/\hbar} \, Ai\left[\Bigl(\frac{3\, V_3 }{8\, \pi \,G\, \hbar\, l } \Bigr )^{2/3} (q_0 + l^2)\right] = \Bigl(\frac{8 \pi \, G \hbar \, l}{3 \, V_3}\Bigr)^{2/3} e^{- \frac{i \,\Pi_0 \, l^2}{3 \,\hbar} \Bigl((\frac{8\, \pi \, G\,l}{3 \,V_3})^2\,\Pi_0^2 + 3 \Bigr)} \,,
\end{align}
where the integral must be performed over all real $q_{0}$, i.e. all the sizes of three-sphere at $r = 0$, including also possible changes in signature\footnote{We want to remind the reader that the path integrals we consider involve in general complexified metrics.}. Upon using the relation~\eqref{momentumcondition} between $\Pi_{0}$ and $\alpha$ and up to an overall normalisation that we were persistently ignoring throughout the text we recognize in the outcome the partition function for the Neumann condition at $r = 0$ given by Eq.~\eqref{eq.ZS3Neumannexact}. Note that this relation applies only in the limit $R_{3} \rightarrow \infty$ and at finite $R_3$ it can only be approximate. \\
Thus, implicitly extending the Dirichlet result to $q_0 \neq 0,$ this relation provides a link between \eqref{eq.ZS3Neumannexact} and \eqref{Dresult} in the limit where $R_3 \to \infty,$ as in that limit the second line in \eqref{Dresult} disappears.
In other words, in this limit the Neumann calculation represents the momentum space wavefunction compared to the position space Dirichlet case as indeed the path integrals \eqref{eq.ZS3Neumannexact} and \eqref{Dresult} satisfy the Wheeler-DeWitt equations in momentum and position space respectively.  \\
These considerations show that the close agreement between the Neumann and Dirichlet results is truly accidental, as conceptually these two calculations are very different. As we will argue below, the associated thermodynamic interpretations must therefore also differ. It will be interesting to understand the holographic interpretation of these two conditions. 

The calculations that we have presented so far have direct analogues in early universe cosmology. Before exploring the implications of this correspondence in section \ref{sec:cosmology}, we will however first deepen our results by considering the addition of black holes with AdS asymptotics as saddles.


\section{$S^2 \times S^1$ boundary and black holes as saddles} \label{sec:bhs}

\subsection{Preliminaries}

In the previous section we saw how to obtain Euclidean AdS$_{4}$ space from a path integral with a fixed three-sphere boundary. In order to include black holes in our discussion, and to see how classic results such as the Hawking-Page phase transition appear in our framework, we must change the topology of the boundary to a direct product of a two-sphere and a circle, as sketched in Fig. \ref{fig:partition}. We will proceed in much the same way as in the previous section, but the added complications of the metric ansatz make us focus on the saddle point approximation and the choice of contour in the underlying gravitational path integral. We will not include counterterms since we will use the empty AdS solution as our reference solution in the partition function, as is often the case in the holographic literature. Also, the question of which conditions should be used on the inner boundary of the integration range is rather subtle, and we will discuss it in detail.

The metric ansatz that we will use in the following is given in Eq.~\eqref{eq.bhmetric} and in the context of minisuperspace approaches appeared earlier starting from the work by Halliwell and Louko~\cite{Halliwell:1990tu}. We have adopted a convention for the lapse function $N$ such that for real $N$ and $b/c > 0$ the coordinate $r$ is timelike. However, in light of the black hole solutions presented in section \ref{sec:metrics}, we should expect the saddle point values of the lapse to turn out imaginary, thus rendering the metric Euclidean. There are two scale factors, $b(r),$ which determines the size of the $S^2,$ and $c(r)$ which determines the size of the Euclidean time direction $\tau.$ Moreover, we will take the $\tau$ direction to be periodically identified with period $\Delta \tau$, such that it will have the topology of a circle. We will once again assume a finite range of the $r$ coordinate, $0 \leq r \leq 1$ with the inner boundary at $r=0$ and the outer boundary at~$r=1.$

Let us immediately discuss the required boundary conditions. On the outer boundary at $r=1,$ we will impose Dirichlet boundary conditions, keeping the proper size of the outer boundary fixed. If we denote the size of the boundary circle by $R_1$ and that of the boundary two-sphere by $R_2,$ then that means that we will impose
\begin{align}
b(r=1) \equiv R_2 \quad \mathrm{and} \quad \sqrt{\frac{c(r=1)}{b(r=1)}}\Delta \tau \equiv R_1\,. \label{Dr1}
\end{align}
Note that both the form of the metric~\eqref{eq.bhmetric} and the above boundary condition are preserved under a residual diffeomorphism and redefinition of functions defining metric components:
\begin{equation}
\tau \rightarrow \gamma \, \tau, \quad c \rightarrow \gamma^{-2} \, c \quad \mathrm{and} \quad N \rightarrow \gamma^{-1} \, N. \label{residual}
\end{equation}
The easiest way to fix this gauge freedom is to fix the periodicity of the $\tau$ coordinate to a convenient value, as we will do below.

In order to obtain a variational problem consistent with Dirichlet boundary conditions as given by Eq.~\eqref{Dr1}, we will have to add the usual GHY term at the outer boundary.
As we will mention in a little more detail below, imposing Dirichlet boundary conditions on the inner boundary leads to results that are inconsistent with the interpretation of the gravitational path integral as the partition function. The Hawking-Page calculation of black hole thermodynamics in asymptotically AdS space in fact assumed that there was no surface term on the inner boundary (which coincides with the horizon location of the black holes) and indeed on shell the geometry smoothly caps off at $r = 0$. This suggests that off-shell we should impose Neumann conditions at $r = 0$, i.e. that we should fix the momenta rather than the field values as we did before in section~\ref{subsec:N}. In fact, we will view not including the GHY term at $r = 0$ and getting a well defined path integral in the minisuperspace approach as a covariant definition of imposing there the Neumann condition\footnote{In dimensions other than four, a surface term is required to obtain a Neumann boundary condition \cite{Krishnan:2016mcj}.}.

To get started, let us evaluate the extrinsic curvature that enters the GHY term for the metric~\eqref{eq.bhmetric}. To this end, at a fixed radius $r$ our ansatz describes a $S^1\times S^2$ manifold with a diagonal metric
\begin{align}
h_{ij} = \mathrm{diag}_{ij} \left(\frac{c}{b},\, b^2,\, b^2 \sin^2 \theta \right)\,.
\end{align}
The conjugate momenta are defined in terms of the extrinsic curvature $K_{ij}$ via
\begin{align}
\Pi^{ij} \equiv -\frac{\sqrt{h}}{16\pi G}\left(K^{ij} - h^{ij}K \right)\,,
\end{align}
which leads to
\begin{align}
\label{eq.PiBH}
\Pi^{ij} = -\frac{1}{16\pi G} \, \mathrm{diag}^{i j} \left[ 2 \, \frac{b\,\dot{b}}{N}, \frac{1}{2}\left(\frac{\dot{c}}{N\,b}+\frac{c\,\dot{b}}{N\,b^2}\right), \frac{1}{2 \sin^2\theta}\left(\frac{\dot{c}}{N\,b}+\frac{c\,\dot{b}}{N\,b^2}\right)\right]\,.
\end{align}
The total GHY surface term is given by the sum of the products of momenta and fields,
\small
\begin{align}
\Pi^{ij}h_{ij} = \frac{1}{8\,\pi \,G}\sqrt{h}\, K = - \frac{1}{16\,\pi\, G}\left( \frac{b\,\dot{c}}{N} + 3\frac{c\,\dot{b}}{N}\right)\,.
\end{align}
\normalsize
Based on these considerations we write the action as 
\begin{align}
S = \frac{1}{16\pi G}\int d^4 x \sqrt{-g}\left[ R + \frac{6}{l^2} \right] + \frac{1}{8\pi G}\int_{outer} d^3 y \sqrt{h} K \,.
\end{align}
With these mixed Neumann conditions at $r=0$ and Dirichlet conditions at $r=1$, the minisuperspace action reduces to
\begin{align}
\label{eq.SND}
S_{ND} = \frac{\Delta \tau}{2G} \int  dr \, \left[ -\frac{\dot{b}\,\dot{c}}{N} +N\left(1 + \frac{3 \, b^2}{l^2} \right)  \right] \, - \, \frac{\Delta\tau}{4 G} \left( \frac{b\,\dot{c}}{N} + 3\frac{c\,\dot{b}}{N}\right)\Bigg|_{r=0}\,.
\end{align}
Varying the action with respect to $b$ and $c$ gives
\begin{align}
\delta S_{ND} & = \frac{\Delta \tau }{2 G} \int dr \left[ \left(\frac{\ddot{c}}{N} + \frac{6 N b }{l^2}\right)\delta b + \left(\frac{\ddot{b}}{N}\right) \delta c \right] \nonumber \\ & - \frac{\Delta \tau }{2 G} \left( \frac{\dot{c_1}}{N} \delta b_1 +  \frac{\dot{b}_1}{N} \delta c_1\right) -\frac{\Delta \tau }{4 G} \left( b_0^2 \delta\left(\frac{\dot{c}_0}{N b_0}\right) + \dot{b}_0 \delta c_0  + 3  c_0 \delta \dot{b}_0  \right)\,.
\end{align}
Let us now explore possible boundary conditions that render the variational problem well-posed. As we anticipated, at $r = 1$ we will impose Dirichlet boundary conditions, which ensure~\eqref{Dr1} and make the respective boundary terms disappear. The situation at $r = 0$ is more subtle. Cancelling the last two terms at $r = 0$ in Eq.~\eqref{eq.SND} can be achieved by setting
\begin{equation}
\label{eq.c0condBH}
c(0) = c_{0} = 0.
\end{equation}
Note that despite the fact that it looks like a Dirichlet condition for $c$, we view it as a Neumann condition from the geometric point of view. Getting rid of the first term at $r = 0$ can be done either with
\begin{equation}
\label{eq.b0condBHve0}
b_{0} = 0 \quad \mathrm{or} \quad \frac{\dot{c}_0 \Delta \tau}{N \, b_0} = \mathrm{fixed}.
\end{equation}
In the following we will encapsulate both conditions in the form of the following single equation\footnote{There is a choice of sign on the right hand side, which is analogous to the choice of sign we encountered with the momentum condition \eqref{momentumcondition} in section \ref{sec:ads}. We choose this sign such that the black hole solutions are dominant over singular saddle points, rather than other way around (cf. the discussion below). This means that we will take $\omega$ to be a positive real number. \label{footomega}}
\begin{equation}
\label{eq.horcond}
\frac{\dot{c}_0 \, \Delta \tau}{4 \, \pi \,i \, N \, b_0} \equiv \omega = \mathrm{fixed}.
\end{equation}
This equation has a simple interpretation when dealing with Euclidean metrics for which $b_{0}, \, \dot{c}_0$ and $i \, N$ are positive, where $\omega$ is related to the deficit / excess angle spanned by the $S^{1}$ direction at $r = 0$. When $b_{0} = 0$, which corresponds to $\omega \rightarrow \infty$, the $S_{1}$ direction does not shrink to a zero size at $r = 0$. This is the case for the thermal AdS solution given by Eqs.~\eqref{eq.defbBH} and~\eqref{eq.BHNc} with $r_+ = 0$ and arbitrary periodicity in $\tau$. For $\omega = 1$ the geometry smoothly ends at $r = 0$ without a conical singularity, as is the case for the Euclidean AdS black hole. For any other value of $\omega$ we end up with a conical singularity at $r = 0$.

With the variational problem well-posed, we can proceed to perform the path integrals over the scale factors. Since the action is again quadratic in $b$ and $c$ we may evaluate the path integrals over these fields in analogy with the integration over $q$ in section \ref{sec:ads}, i.e. by shifting the variable of integration to the sum of a solution of the equations of motion plus a general fluctuation~\cite{Halliwell:1990tu}. The fluctuation integrals will be unimportant, since they will lead to an overall numerical prefactor in front of the partition function, see appendix \ref{sec:determinant}. The nontrivial physics lies in the solutions of the equations of motion for $b$ and $c,$ given by
\begin{align}
\label{eq.eomsthermal}
\ddot{b} = 0 \quad \mathrm{and} \quad \ddot{c} = -\frac{6}{l^2} N^2 b\,.
\end{align}
The solutions of Eqs.~\eqref{eq.eomsthermal} subject to the conditions~\eqref{Dr1}, \eqref{eq.c0condBH} and~\eqref{eq.horcond} take the form
\begin{subequations}
\begin{align}
b(r)  &= (R_{2}-b_0)\,r+b_0 \,,\label{bsol}\\
c(r)  &= \left( c_1 + \frac{(2 \, b_{0}+ R_{2})N^2}{l^2} \right) r - \frac{3 \, b_{0} N^2}{l^2} \, r^2 + \frac{(b_{0} - R_{2})N^2}{l^2} r^3\,, \label{csol}
\end{align}
\end{subequations}
where
\begin{equation}
b_{0} = \frac{\Delta\tau \left(l^2 \, c_{1} + N^2 \, R_2\right)}{4 \,\pi \, i \, \omega \, l^2 \, N - 2 \Delta\tau \, N^2}\,, \qquad  c_1 = \frac{R_1^2 R_2}{\Delta\tau^2}\,.
\end{equation}
With these solutions at hand, we may now perform the integrations over $b$ and $c,$ leaving us with an integral for the lapse function only. 

\subsection{Evaluation of the gravitational path integral}

The partition function once again reduces to an ordinary integral over the lapse function, with two additional features: first, we must include a suitable integral over the boundary conditions on the inner boundary, i.e. we must include a sum over $\omega$, and secondly, we will implement a background subtraction and use the AdS solution given by Eqs.~\eqref{eq.defbBH} and~\eqref{eq.BHNc} with $r_+ = 0$ as a reference. Thus the partition function is given by
\begin{align}
Z(R_1,R_2) = \int d\omega\int dN e^{\frac{i}{\hbar}\left(S_{ND}(N) - S_{EAdS}\right)}
\end{align} 
with
\begin{align}
&\frac{8\,G \, l^2}{\Delta\tau}S_{ND} = \nonumber \\ &\frac{(3 R_2^2 + 4 l^2)\Delta\tau N^4 - 8 \pi i \omega l^2 (R_2^2 + l^2) N^3 - 6 l^2 R_2 c_1 \Delta \tau N^2 + 8 \pi i \omega l^4 R_2 c_1 N - l^4 c_1^2 \Delta \tau}{N^2 (\Delta \tau N -2\pi i \omega l^2)}\,. \label{lapseaction}
\end{align}
The partition function is a sum over interior geometries with the boundary at $r=1$ held fixed. When searching for saddle points, the only smooth geometries correspond to $\omega = 1$ (black holes) or the limit $\omega \rightarrow \infty$ (thermal AdS). We restrict ourselves to geometries that on-shell are smooth at $r = 0$, in which case the integral over $\omega$ thus reduces to a sum over just two values,
\begin{align}
  Z(R_1,R_2) =  \sum_{\omega = 1, \infty} \int dN e^{\frac{i}{\hbar}\left(S_{ND}(N) - S_{EAdS}\right)}\,.
\end{align}
It would be very interesting to include geometries with conical deficits in our framework and we leave it for future work.

We will analyse the full saddle point structure momentarily, but first we may check explicitly that the black hole and AdS solutions will arise from this action. It is not possible to solve for the saddle points of the action analytically, since the corresponding equation $\frac{d S_{ND}}{d N}=0$ is a quintic. There are generally five distinct saddle points. However one may verify by direct substitution that two of them are given by
\begin{subequations}
\begin{align}
\label{cbh}
N_s & = -i (R_2 - r_+)\,, \\ b(r) &= r (R_2-r_+) + r_+\,, \\ c(r)  &= \frac{1}{l^2}[b^3(r) + l^2 b(r) - r_+^3 - l^2 r_+],.
\end{align}
\end{subequations}
with $r_+$ taking the two possible values that solve \eqref{horizon} for a given $\beta,$ and where we have fixed the scaling ambiguity \eqref{residual} by choosing
\begin{equation}
\Delta \tau^{bh} = \beta.
\end{equation} 
These solutions satisfy the boundary condition $\omega=1.$ Evidently these are the sought after black hole solutions presented in section \ref{sec:metrics}. Their action is given by
\begin{align}
S^{bh} = -\frac{i}{4Gl}\sqrt{\frac{R_1^2 R_2}{(R_2 - r_+)(R_2^2 + l^2 + R_2 r_+ + r_+^2)}} \left(4 R_2 (R_2^2 + l^2) - 3 l^2 r_+ - r_+^3 \right)\,. \label{actionbh}
\end{align}
Note that the black hole solutions only arise on the negative imaginary lapse action. This is because our boundary condition \eqref{eq.horcond} has broken the invariance of the action under complex conjugation, cf. also footnote \ref{footomega}. 

We expect to recover the pure AdS solution as the limit where $\omega \to \infty.$ In this limit the action reduces to
\begin{align}
S_{ND,\omega \to \infty} = \frac{\Delta \tau}{2G l^2}\left( (R_2^2 + l^2) N - \frac{R_2 c_1 l^2}{N} \right)\,.
\end{align}
The saddle points are found to correspond to EAdS space, as expected, with 
\begin{align}
N_s =  \pm i R_2\,, \quad b(r) = R_2 r\,, \quad \frac{c(r)}{b(r)} = 1 + \frac{R_2^2 r^2}{l^2}\,.
\end{align}
The scaling ambiguity \eqref{residual} has been fixed in such a way that the AdS solution corresponds to the limit $r_+ \to 0$ of the black hole solution above. The periodicity in the $\tau$ direction must then be chosen such that the circle size on the outer boundary remains $R_1,$ namely
\begin{align}
\Delta \tau^{AdS} = \frac{R_1 l}{\sqrt{R_2^2 + l^2}}\,.
\end{align}
The action for these saddle points is
\begin{align}
S^{AdS} = \pm i \frac{R_1 R_2}{G l}\sqrt{R_2^2 + l^2}\,. \label{actionads}
\end{align}

\begin{figure}
\includegraphics[width=0.4\textwidth]{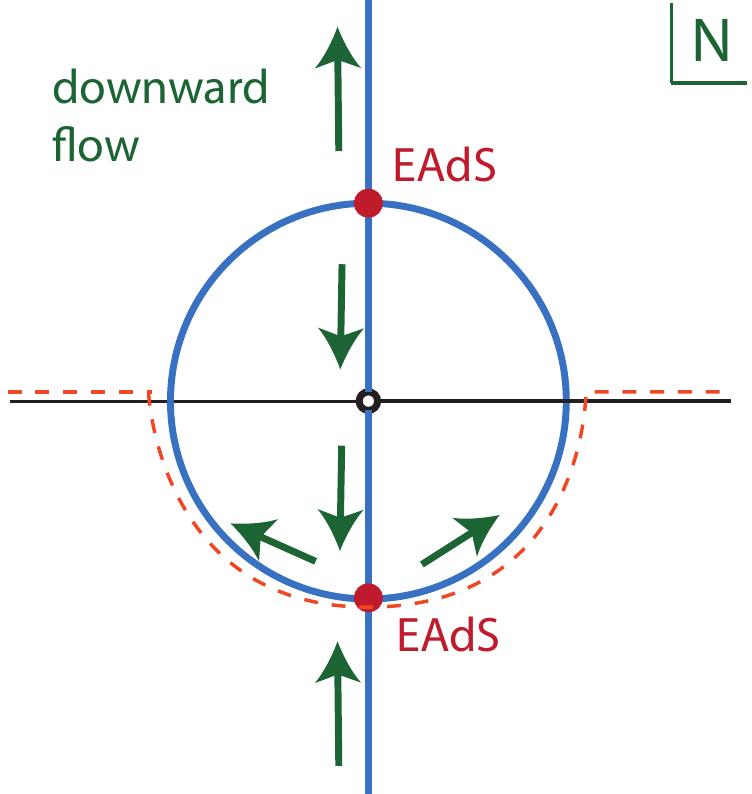}
\caption{Flows in the AdS case, which corresponds to the limit $\omega \to \infty.$ Arrows indicate the direction of steepest descent from the two saddle points. There is a singularity at $N=0.$ The dashed line indicates the required contour of integration which picks up a contribution from the enhanced EAdS saddle point in the lower half plane.} \label{fig:ads2}
\end{figure} 

The question now is which saddle points contribute? And what do the additional saddle points represent? We will look at the saddle point structure, and the associated paths of steepest descent, numerically. 

We start with the limiting case where $\omega \to \infty.$ As we have just derived, there are two saddle points in this case, which are complex conjugates of each other. They both describe EAdS space, but with different weightings. The corresponding flow lines are shown in Fig.~\ref{fig:ads2}, where the steepest descent paths emanating from the saddle points are drawn. The saddle point with the enhanced weighting is the one in the lower half plane. As in the case of the three-spheres partition function with Dirichlet boundary conditions, there is a singularity at $N=0.$ We may interpret the singularity in the action as a signal that the corresponding geometry does not exist. A suitable contour of integration is indicated by the dashed line in the figure. We must choose it such that at large $|N|$ it resides asymptotically in the upper half plane (for convergence) and such that it passes below the singularity so as to pick up a contribution from the enhanced EAdS saddle point. It would have been possible to define a Euclidean contour along the positive imaginary lapse axis, but such a contour would only have picked up the suppressed EAdS saddle point, which disappears in the limit of a large boundary size. Thus we are again forced to integrate over complex metrics.  

\begin{figure}
\includegraphics[width=0.5\textwidth]{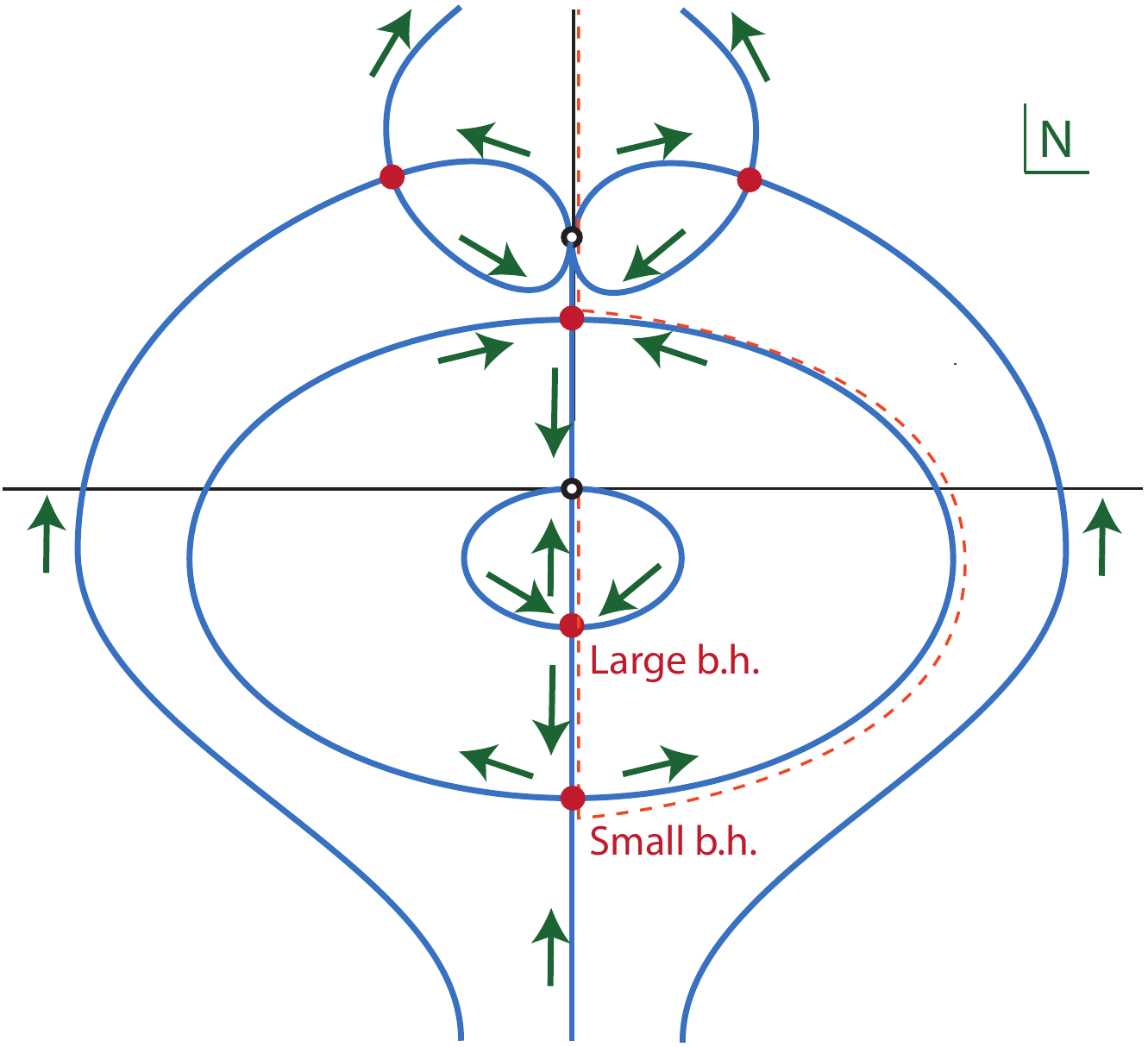}
\caption{Saddle points (in red) and steepest descent lines (in blue) for $R_1=5, R_2 = 10, l=1.$ Arrows indicate directions of steepest descent. There are two singularities, one at $N=0$ and one on the positive imaginary $N$ axis. In dashed orange is the required contour of integration.} \label{fig:bh1}
\end{figure} 

\begin{figure}
\includegraphics[width=0.3\textwidth]{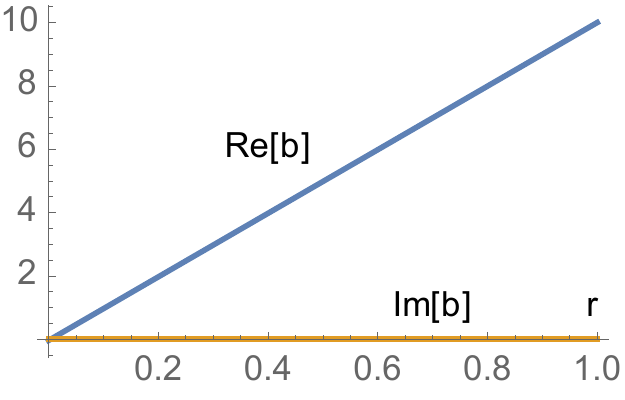}
\includegraphics[width=0.3\textwidth]{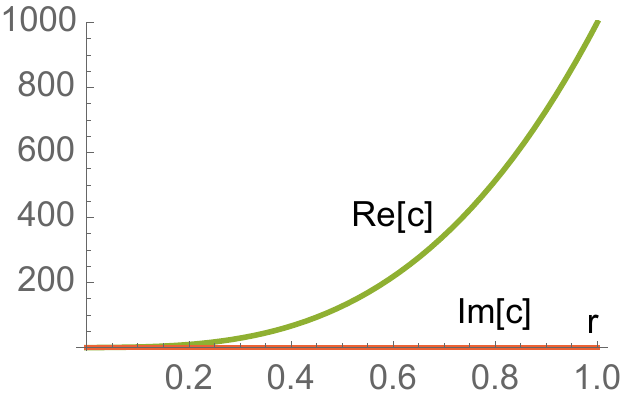}
\caption{The geometry of the Euclidean saddle point on the positive imaginary axis. Both $b(r)$ and $c(r)$ are real valued. At the origin $b$ has a small negative value and then passes through zero to reach the final value $b(1)=10.$ Here we used the same parameter values as in Fig. \ref{fig:bh1}.} \label{fig:bh1sad1}
\end{figure}

\begin{figure}
\includegraphics[width=0.3\textwidth]{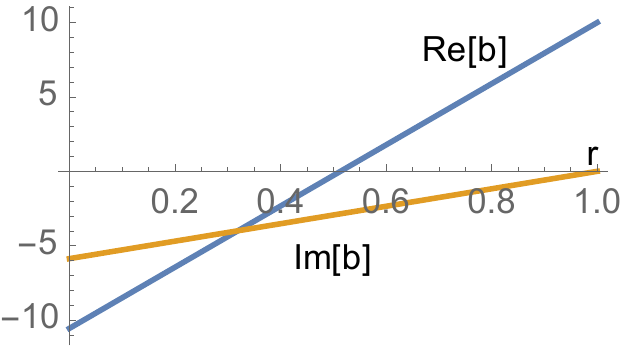}
\includegraphics[width=0.3\textwidth]{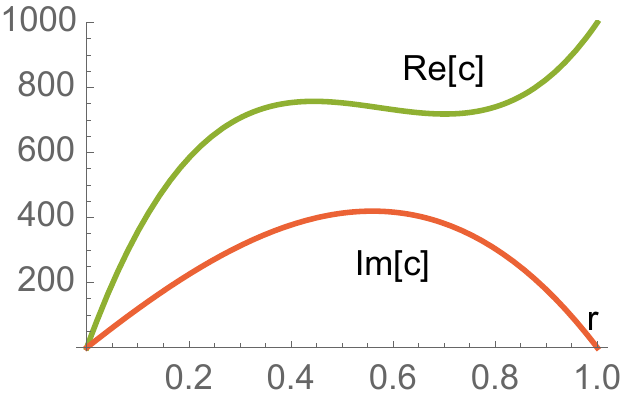}
\caption{The geometry of the complex saddle point in the first quadrant of Fig. \ref{fig:bh1}. Both $b(r)$ and $c(r)$ are complex valued. Here we used the same parameter values as in Fig. \ref{fig:bh1}.} \label{fig:bh1sad2}
\end{figure}

In the following figures, we will analyse the contributions from $\omega=1,$ at a fixed two-sphere radius $R_{2}$ and for increasing circle radii $R_1.$ The case of having a very small $R_1$ is shown in Fig. \ref{fig:bh1}. Three additional saddle points appear here: one near the origin, and two complex conjugate saddle points that move in from infinity (as $R_1$ is increased from zero) in the upper half plane. We will briefly describe the saddle points, starting with the two that already existed when $\omega \to \infty,$ i.e. what were the enhanced EAdS and the suppressed EAdS solutions. The enhanced EAdS solution now turns into the small Euclidean black hole, with horizon size $r_+$ growing from zero as $R_1$ is increased. Meanwhile, what was the suppressed EAdS solution turns into a geometry that starts out at a (small) negative value of $b,$ barely visible in Fig. \ref{fig:bh1sad1}. This means that the metric signature is $(-,-,+,+)$ near the origin, and then turns Euclidean after $b$ has crossed zero. Since $b$ crosses zero we may expect perturbations to blow up at that location. The most important saddle point is the one that appeared near the origin on the negative imaginary axis. This is the large black hole, with $r_+$ corresponding to the larger solution to \eqref{horizon}. This is the dominant saddle point, with the highest weighting. In fact the steepest descent line from this saddle point moves down towards the singularity, and on the other side of the saddle point down towards the small black hole and from there on to the Euclidean saddle point in the upper half plane. The two remaining saddle points in the upper half plane have a suppressed weighting, and their geometry is shown in Fig. \ref{fig:bh1sad2}. Their metric is complex throughout, with the exception of the final boundary at $r=1.$ A suitable contour of integration, capturing the large black hole, is indicated by the dashed line in Fig.~\ref{fig:bh1}. It has different characteristics than the one in the infinite $\omega$ case, namely it emanates from the origin in the negative imaginary direction, then winds around the singularity and ends up shooting off to infinity in the upper half plane. The contour has to start at the singularity in the direction of the lower half plane in order to capture the large black hole. This is possible because at finite $\omega$ the lapse integrand \eqref{lapseaction} near $N=0$ behaves as $e^{+\frac{1}{N^2}},$ implying that there exists a region of convergence in the wedge surrounding the imaginary axis at $\pm 45 ^o.$ A purely Euclidean contour is however not possible since the asymptotic region at large negative imaginary values of the lapse is a region of divergence, and hence the contour must wind around towards the upper half plane. Despite the fact that the contour differs from the infinite $\omega$ case, a similarity is that we are once again forced to integrate over complex metrics in order to obtain sensible results. 

\begin{figure}
\includegraphics[width=0.5\textwidth]{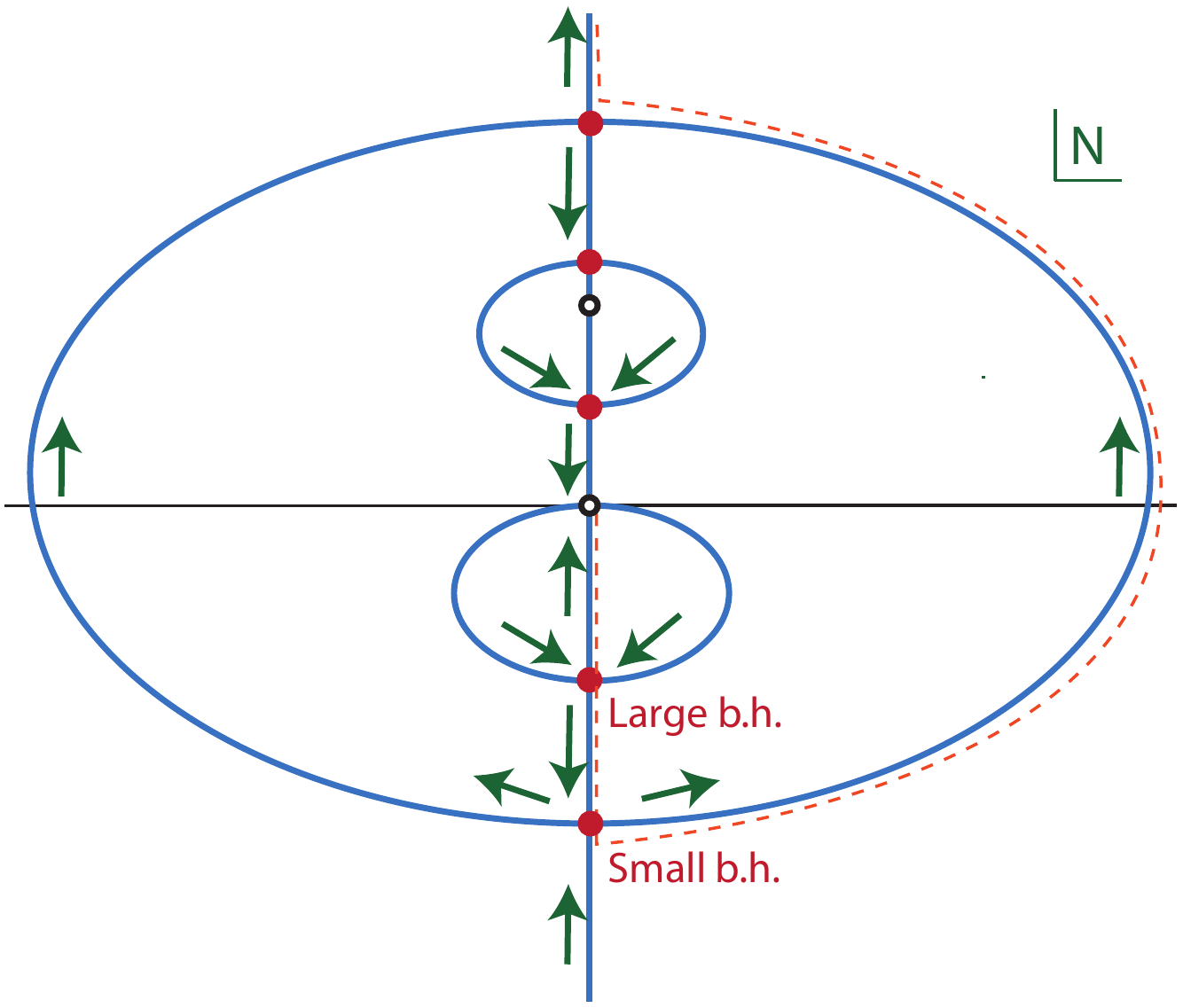}
\caption{Saddle points (in red) and steepest descent lines (in blue) for $R_1=12, R_2 = 10, l=1.$ Arrows indicate directions of steepest descent. There are two singularities, one at $N=0$ and one on the positive imaginary $N$ axis. In dashed orange is the required contour of integration.} \label{fig:bh2}
\end{figure}

\begin{figure}
\includegraphics[width=0.5\textwidth]{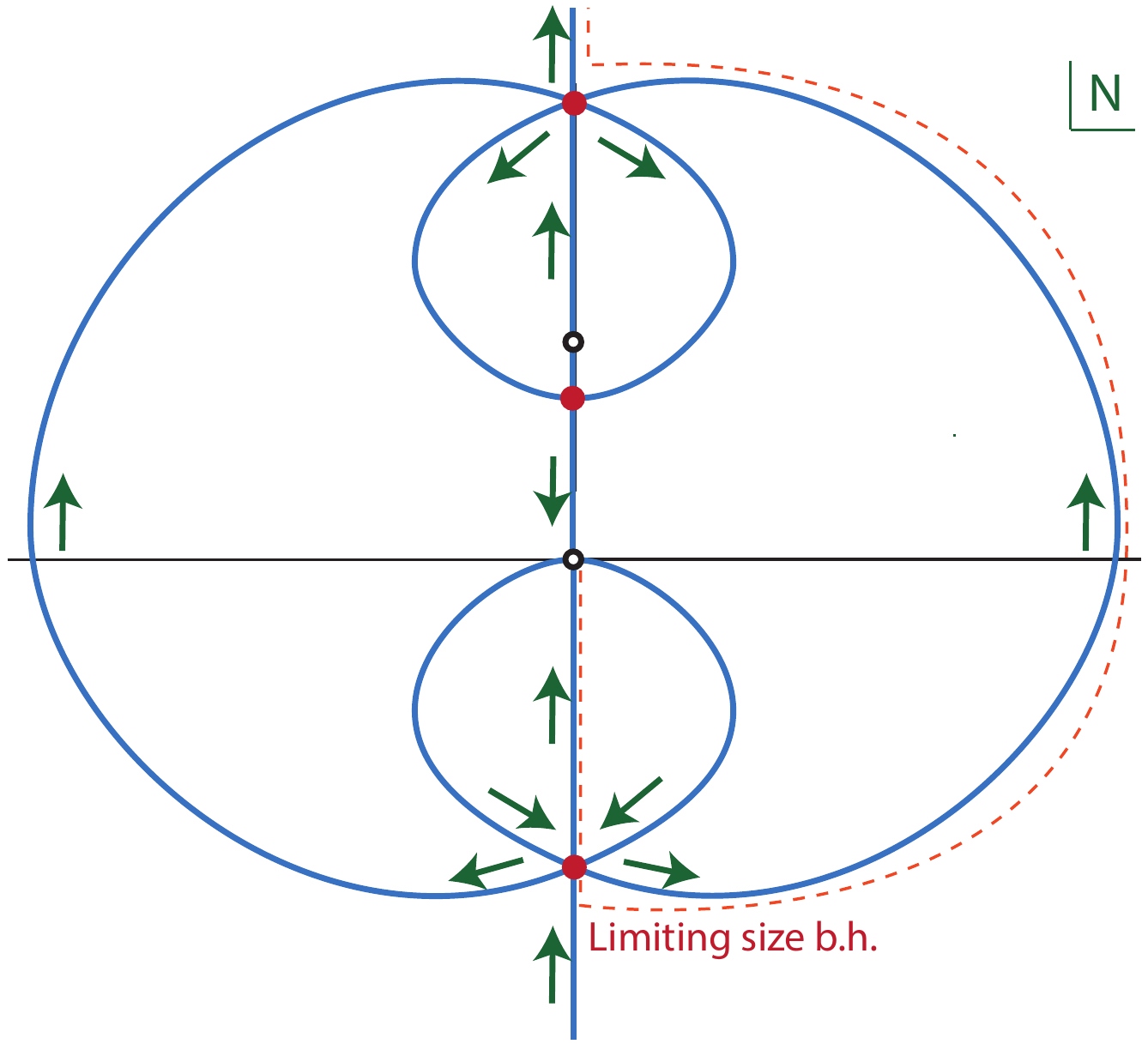}
\caption{Saddle points (in red) and steepest descent lines (in blue) for the limiting case $R_1= R_{1,limit}, R_2 = 10, l=1.$ Arrows indicate directions of steepest descent. In dashed orange is the required contour of integration.} \label{fig:bh3}
\end{figure}

\begin{figure}
\includegraphics[width=0.3\textwidth]{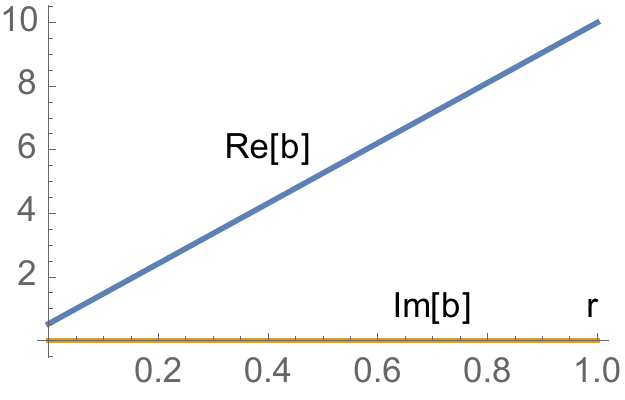}
\includegraphics[width=0.3\textwidth]{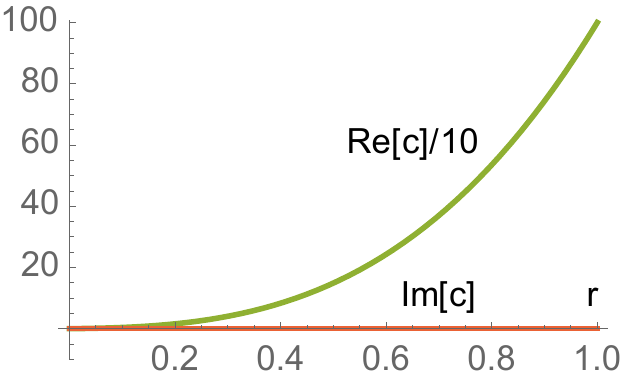}
\caption{The geometry of the limiting black hole, i.e. for the case where the small and large black holes have merged. Both $b(r)$ and $c(r)$ are real valued, and the saddle point geometry is Euclidean. Here we used the same parameter values as in Fig. \ref{fig:bh3}.} \label{fig:bh3sad}
\end{figure}

\begin{figure}
\includegraphics[width=0.5\textwidth]{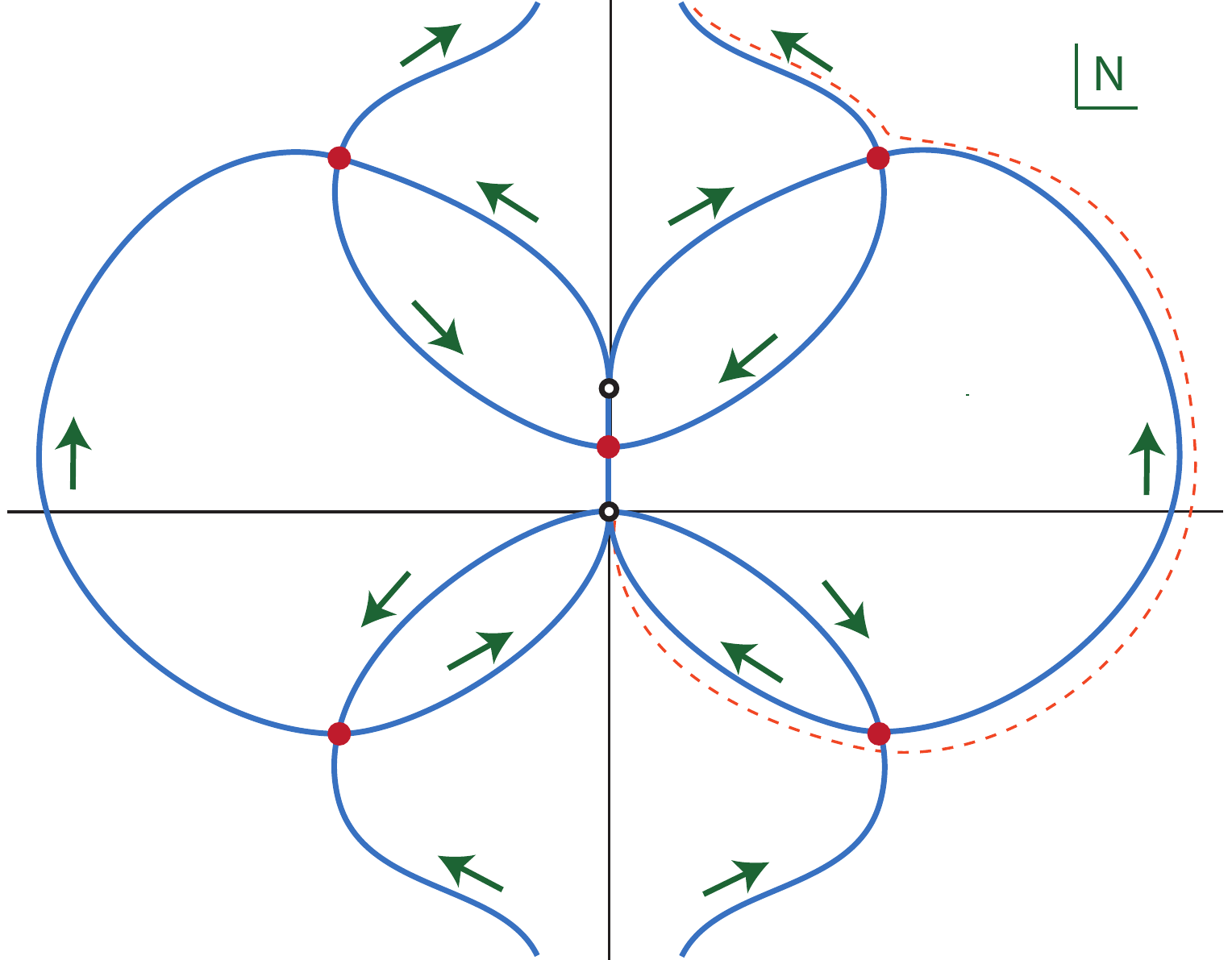}
\caption{Saddle points (in red) and steepest descent lines (in blue) for $R_1=50, R_2 = 10, l=1.$ Arrows indicate directions of steepest descent. There are two singularities, one at $N=0$ and one on the positive imaginary $N$ axis. In dashed orange is the required contour of integration.} \label{fig:bh4}
\end{figure}

\begin{figure}
\includegraphics[width=0.3\textwidth]{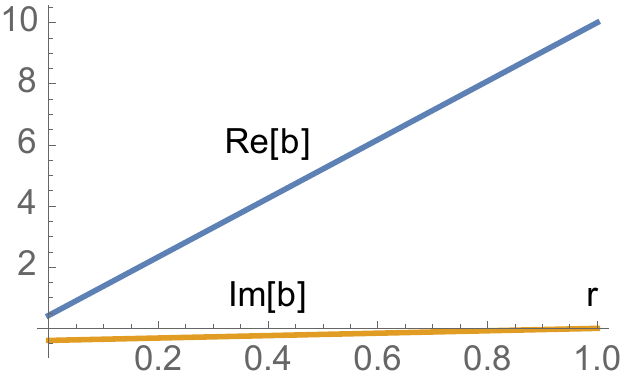}
\includegraphics[width=0.3\textwidth]{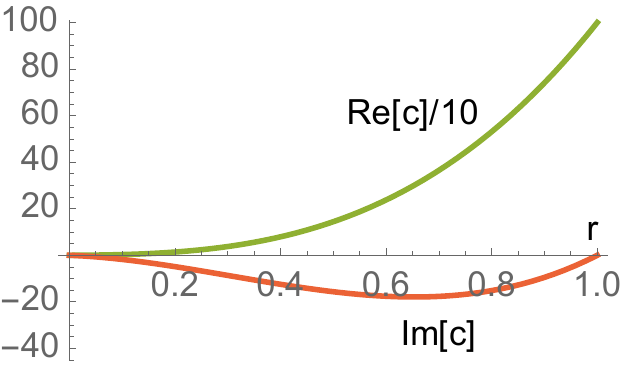}
\caption{At low temperature, when $R_1$ is large, the saddle points that used to correspond to black holes have moved into the complex plane. The associated geometry is no longer Euclidean, as imaginary parts of $b(r)$ and $c(r)$ develop. Here we used the values $R_1 = 50, R_2 = 10, l=1,$ just as in Fig. \ref{fig:bh4}.} \label{fig:bh4sad}
\end{figure}

As $R_1$ is increased, there are few relevant changes at first. The complex saddle points move towards the Euclidean axis, merge there and then separate again into two further Euclidean solutions similar to the one shown in Fig.~\ref{fig:bh1sad1}. The saddle points and their flow lines are shown in Fig.~\ref{fig:bh2}. The most important change occurs once $R_1$ reaches the limiting value $R_{1,limit}$ -- this case is shown in Fig.~\ref{fig:bh3}. At this radius, the two black hole solutions merge into a degenerate saddle point, representing the black hole at the minimum temperature (maximum radius) that is required for black holes to exist. This limiting black hole geometry is shown in Fig.~\ref{fig:bh3sad}. One may obtain an expression for the limiting radius by combining Eqs. \eqref{horizon}, \eqref{Dr1} and \eqref{cbh}, with $\Delta\tau=\beta$ and inserting the maximum value for the periodicity $\beta_{max} = \frac{2 \pi l}{\sqrt3}$, which leads to 
\begin{align}
\label{eq.R1limit}
R_{1,limit} = \frac{2\pi}{\sqrt{3}}R_2 \sqrt{1+ \frac{l^2}{R_2^2}- \frac{4l^3}{3\sqrt{3}R_2^3}}\,.
\end{align}
Once the circle radius $R_1$ is increased even further, the degenerate black hole saddle point, as well as two of the saddle points in the upper half plane, all move into the complex plane -- see Fig. \ref{fig:bh4} for a depiction of the saddle point locations and the associated steepest descent flows. At that stage there does not exist any Euclidean black hole solution anymore. The complex saddle points possess a fully complex geometry, which is shown in Fig. \ref{fig:bh4sad}. In our framework, this is the manifestation of the well-known fact that there exists a minimum temperature required for the existence of regular Euclidean black holes. The complex saddle points have a suppressed weighting, smaller in fact than the empty EAdS solution, as we will show below. 

In all cases one is forced to choose a contour of integration that starts at the singularity at $N=0,$ then follows the thimble associated with the large black hole saddle point, curves around and eventually flies off to infinity in the upper half plane, as required for convergence. Note that once again the integrand possesses the symmetry that for $N \rightarrow - N^*$ it changes to its complex conjugate. Thus in order to obtain a real partition function picked around the black hole saddle points one should consider the contour described above together with its reflection with respect to the imaginary lapse axis. As we commented on already for the case of the $S^3$ boundary, this symmetric contour is as close as it can get to the integral along the imaginary axis i.e. the Euclidean path integral, which in itself is divergent and thus ill-defined. The interesting point is that this contour neither corresponds to a sum over Euclidean metrics, nor a sum over Lorentzian metrics -- in order for the partition function to be mathematically meaningful as a minisuperspace statement, the sum must be defined over intrinsically complex metrics.

\begin{figure}
\includegraphics[width=0.5\textwidth]{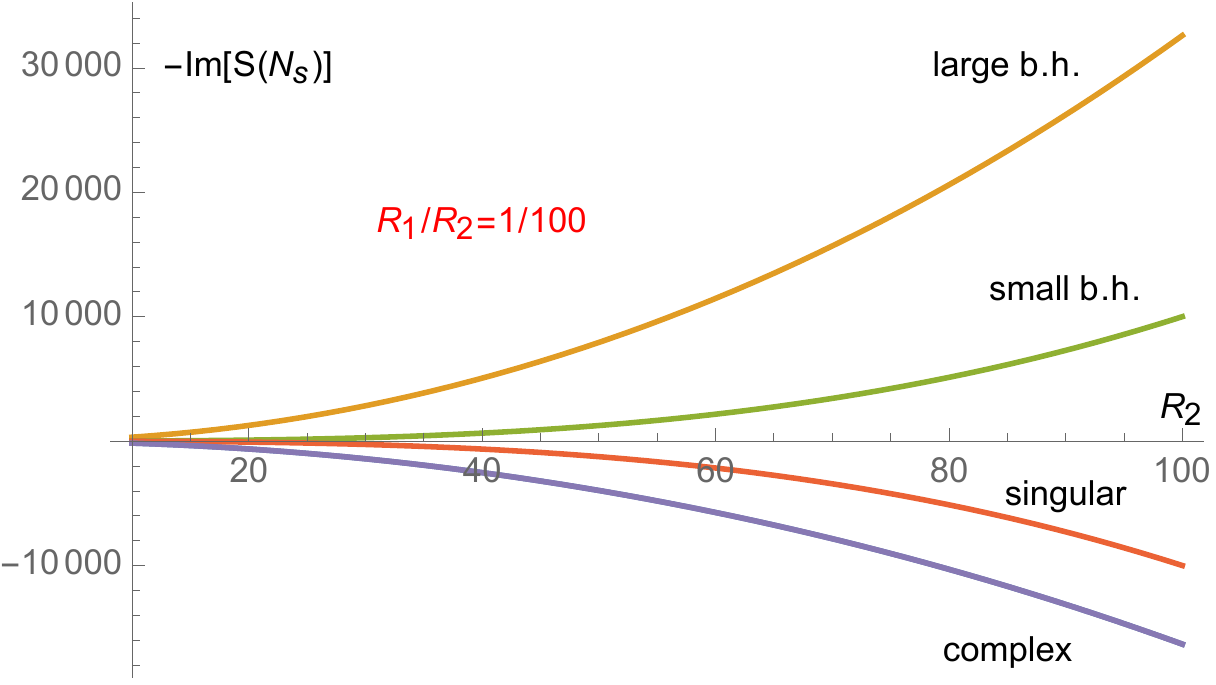}
\caption{Weighting of the saddle points $- Im[S(N_s)]$ with constant ratio $R_1/R_2=1/100,$ as a function of the boundary size and with $8 \, \pi \, G = 1$ and $l=1$. One can see that in the limit of infinite boundary size only the black hole saddle points will remain.} \label{fig:ratio}
\end{figure} 

\subsection{Thermodynamics from saddles}

Having discussed the saddle points and integration contours, we may now sketch how the usual  interpretation in terms of thermodynamics is recovered. In all cases we saw that the contribution to the partition function from $\omega = 1$ is dominated by the large black hole solution, provided $R_1$ is smaller than the limiting value~\eqref{eq.R1limit}. Thus, when approximating the partition function, the $\omega = 1$ contribution may be well approximated by the action of the large black hole solution. The difference in action between the black hole and AdS solution is given by the difference between Eq.~\eqref{actionbh} and Eq.~\eqref{actionads}, and as an expansion at large two-sphere radius $R_2$ is given by 
\begin{subequations}
\label{actiondiff}
\begin{align}
\Delta S & = -\frac{iR_1}{4Gl}\left[\frac{\sqrt{R_2}\left(4 R_2^3 +4 l^2 R_2  - 3 l^2 r_+ - r_+^3 \right)}{\sqrt{R_2^3 + l^2 R_2 - l^2 r_+- r_+^3}}   -4 R_2\sqrt{R_2^2 + l^2}\right]\\
&= \frac{i R_1}{4Gl R_2}(l^2 r_+ - r_+^3) -  \frac{i l R_1}{8G R_2^3}(l^2 r_+ - r_+^3)  + O(R_2^{-4})\,.
\end{align}
\end{subequations}
At leading order in a large $R_2$ expansion we may identify $R_1/R_2 \approx \beta/l,$ and with this substitution the leading order difference in actions at large $R_2$ recovers the classic Hawking-Page result~\cite{Hawking:1982dh}
\begin{align}
\Delta S_{HP} = -\frac{i \pi}{G}r_+^2\frac{r_+^2 - l^2}{3r_+^2 + l^2}  + O(R_2^{-1})\,.
\end{align}
The weighting of the AdS solution dominates when $-Im[\Delta S]<0.$ At large $R_2$ this is when $r_+< l,$ and there are corrections, implied by \eqref{actiondiff}, to this relation when $R_2$ is small. The phase transition thus occurs at the approximate radius $R_{1,HP} \approx \pi R_2.$ Thus the complex saddle points that replace the large black hole solutions at large $R_1>R_{1,limit} \approx  \frac{2\pi}{\sqrt{3}}R_2$ never play a dominant role, since the AdS solution has already become dominant by then.

The thermodynamic interpretation follows from an analysis of the partition function, approximated here by the difference in actions \eqref{actiondiff},
\begin{align}
\ln Z = i \frac{ \Delta S}{\hbar}\,.
\end{align}
It is important to keep in mind that we are considering the partition function as representing the canonical ensemble, i.e. we are considering a system that is kept at a fixed temperature~$T.$ At fixed boundary two-sphere with radius $R_2$ this temperature, which is redshifted as one moves away from the black hole horizon, is given by 
\begin{align}
R_1 =\beta \sqrt{1+ \frac{R_2^2}{l^2}-\frac{2M}{R_2}} = \Delta\tau^{AdS} \sqrt{1+ \frac{R_2^2}{l^2}} = \frac{1}{T}\,,
\end{align} 
where we denoted the Euclidean time periodicity of the $EAdS$ solution by $\Delta\tau^{AdS}.$ Thus, reintroducing the speed of light $c,$ we may usefully rewrite the partition function as 
\begin{align}
\ln Z = \frac{R_2}{T \emph{l}^2_P}\left(\sqrt{1+ \frac{R_2^2}{l^2}-\frac{2M}{R_2}} - \sqrt{1+ \frac{R_2^2}{l^2}} \right) + \frac{\pi r_+^2}{\emph{l}^2_P}\,.
\end{align}
where $\emph{l}_P = \sqrt{\frac{G \hbar}{c^3}}$ is the Planck length.\\
The expectation value of the energy is given by
\begin{subequations}
\label{energy}
\begin{align}
\langle E \rangle &=k_B T^2 \frac{\partial \ln Z}{\partial T} = \frac{k_B R_2}{\emph{l}^2_P} \left(\sqrt{1+ \frac{R_2^2}{l^2}} - \sqrt{1+ \frac{R_2^2}{l^2}-\frac{2M}{R_2}} \right) \\
&= \frac{ k_B }{ \emph{l}^2_P}\frac{ l M}{R_2} - \frac{k_B }{\emph{l}^2_P}\frac{ M l^3}{2 R_2^3}+  O(R_2^{-4}) 
\end{align}
\end{subequations}
and the entropy takes the form
\begin{align}
{\cal S} & =k_B \ln Z + \frac{\langle E \rangle}{T} =  \frac{k_B}{\emph{l}^2_P} \, \pi r_+^2 = \frac{k_B}{\emph{l}^2_P} \, \frac{Area}{4} \label{entropy}\,.
\end{align}
Note that this explicitly verifies the Quantum Statistical Relation \cite{Gibbons:1976ue}
\begin{align}
- k_B T \ln Z = \langle E \rangle - T{\cal S}\,.
\end{align}
Furthermore, the results derived above are in agreement with the Hamiltonian method employed by Brown et al. in Ref.~\cite{Brown:1994gs}. In deriving the energy \eqref{energy}, one may use the chain rule that $\partial \ln Z/\partial T = \partial \ln Z / \partial r_+ \left(\partial T/\partial r_+ \right)^{-1},$ with $M$ being thought of as a function of $r_+$ according to \eqref{mass}. The conserved mass differs from the energy by a factor of the lapse at $R_2$
\begin{align}
{\cal M} = \sqrt{1+ \frac{R_2^2}{l^2}-\frac{2M}{R_2}} \langle E \rangle = M + \frac{l^2}{2R_2^2}M - \frac{l^2}{R_2^3}M^2+ O(R_2^{-4})\,.
\end{align}
Note also that the entropy \eqref{entropy} is given precisely by a quarter of the horizon area, and that there are no corrections to this relation at finite $R_2.$ The specific heat at fixed boundary
\begin{align}
{\cal C} = \frac{\partial \langle E \rangle}{\partial T}
\end{align}
is negative for small black holes ($r_+ < \frac{l}{\sqrt{3}}$) and positive for large black holes ($r_+ > \frac{l}{\sqrt{3}}$), which implies that only large black holes are thermodynamically stable. This fits well with our flow diagrams which demonstrate that the large black hole always has a higher weighting than the small black hole, and is thus also more dominant in the canonical ensemble. 

One surprising aspect of our work is the appearance of additional saddle points. These have a weighting that is suppressed compared to the black hole saddle points, both large and small. Thus they do not play a large role. In fact, if the outer boundary is moved all the way to infinity, these additional saddle points disappear altogether -- see Fig. \ref{fig:ratio}. This implies that these extra saddles do not play any role in the original AdS/CFT correspondence and our study reproduces the results of Refs.~\cite{Hawking:1982dh,Witten:1998zw}. However, once the boundary is moved to a finite radius, they will provide a tiny additional contribution to the partition function. It would be very interesting to try to figure out if one can use the appropriate QFT description~\cite{McGough:2016lol,Hartman:2018tkw} to confirm their existence, or rule out the minisuperspace approach.

We would like to emphasise that our choice of mixed boundary conditions is crucial in obtaining an expression for the canonical ensemble. The Dirichlet condition on the outer boundary, which fixes the size of the Euclidean time circle, effectively fixes the temperature. However, this turns out not to be enough. Our premise that we wanted to sum over saddle point geometries that cap off smoothly in the interior led us to the Neumann boundary condition at the coordinate location $r=0.$ Had we used a Dirichlet condition on the inner boundary, i.e. at the black hole horizon, we would have obtained an additional boundary term of magnitude $\pi r_+^2$ contributing to the black hole action. Our path integral would then have been approximated by 
\begin{align}
    -k_B \ln Z \approx \frac{\langle E \rangle}{T} - {\cal S} + {\cal S} = \frac{\langle E \rangle}{T}\,,
\end{align}
which is reminiscent of the discussion in Ref.~\cite{deBoer:2015ija}. Thus, with Dirichlet conditions on both ends, the ``partition function'' would rather have looked like that of the microcanonical ensemble, where one sums over states of fixed internal energy. This is surprising, as in gravitational systems the energy is described by the asymptotic fall-off of the metric, and thus one would have expected the microcanonical ensemble to be given by a path integral with boundary conditions (at the outer boundary $r=1$) that are of Neumann form, where derivatives of the metric may be specified \cite{Brown:1992bq,Krishnan:2016mcj}. More precisely, one might have expected that the canonical and microcanonical ensembles would be related by a Legendre transform at the outer boundary, and not at the inner boundary. The extent to which this correspondence is accidental deserves further investigation.


\section{Implications for cosmology} \label{sec:cosmology}

The no-boundary proposal can be formulated as a path integral in a very similar fashion to the calculations presented in this work (this analogy should already be obvious by taking another look at Fig. \ref{fig:partition}, but rotating the figure by $90$ degrees counter-clockwise). In this context the path integral defines the wave function of the universe which sets the initial conditions for the universe. According to the no boundary proposal the wave function is peaked around a smooth semiclassical geometry where the big bang is replaced by a Euclidean regular section and cosmological fluctuations are suppressed~\cite{Hartle:1983ai,Halliwell:1984eu}. 

The literature on no-boundary path integrals has a long history (see for example \cite{Louko:1988bk}, \cite{Halliwell:1988ik}, \cite{Halliwell:1989dy}, \cite{Halliwell:1990tu}, \cite{Feldbrugge:2017kzv}, \cite{DiTucci:2019bui}). In this section we will review and elaborate on some of the well known results while highlighting the connections with the original calculations of the previous sections regarding the case of a negative cosmological constant. Our aim is to see what one may learn about the no-boundary proposal when viewed from the fresh perspective offered by calculations performed with AdS asymptotics.

When the cosmological constant is positive, $\Lambda = 3\,H^2,$ the classical de Sitter solution with spatial sections that are three-spheres is given by 
\begin{align}
\label{eq.dS}
ds^2 = - dt^2 + \frac{1}{H^2}\cosh^2 (Ht) d\Omega_3^2\,.
\end{align}
Meanwhile, the Euclidean version of this solution is a four-sphere,
\begin{align}
\label{eq.S4}
ds^2 = d\eta^2 + \frac{1}{H^2} \sin^2 (H\eta) d\Omega_3^2\,, \quad t = - i \left(\eta - \frac{\pi}{2H}\right)\,.
\end{align}
Hartle and Hawking's idea was the consider a geometry where the Lorentzian hyperboloid is glued at $t=0$ to half of the Euclidean four-sphere \cite{Hartle:1983ai}. In this way the late time Lorentzian spacetime well approximates an inflationary universe which contains no big bang singularity in the past.

The evaluation of the no-boundary path integral is in one-to-one correspondence with the calculations in section \ref{sec:ads} if one analytically continues the radius of curvature of AdS to an imaginary value $l = \frac{i}{H},$ where $H$ then denotes the Hubble rate of the corresponding de Sitter spacetime. Thus, in our coordinates, the correspondence is simply
\begin{align}
\Lambda = - \frac{3}{l^2} = + 3 H^2. \label{anacon}
\end{align}
It is now interesting to observe that in the case of a negative cosmological constant, evaluating the canonical ensemble in the black hole case required us to impose Neumann boundary conditions on the inner boundary. Not imposing a boundary term resonates well with the philosophy of the no-boundary proposal, as the name itself suggests. The exact same condition, given here by Eq. \eqref{momentumcondition}, was also recently studied in cosmology in Ref.~\cite{DiTucci:2019bui}. Upon performing the analytic continuation in \eqref{anacon}, the saddle points \eqref{S3saddles} come to reside at 
\begin{align}
N_\pm = -\frac{i}{H^2}  \pm \frac{1}{H^2}\sqrt{H^2 R_3^2 - 1}\,,
\end{align}
with associated metrics (compare to Eq. \eqref{saddlemetric})
\begin{align}
\label{eq.nbgeometry}
    ds^2 = -\frac{N_{\pm}^2}{q(r)}dr^2 + q(r)d\Omega_3^2\,, \qquad q(r)=H^2 N_{\pm}^2r^2 + 2N_{\pm} i r\,.
\end{align}
While the saddle points were purely Euclidean with negative cosmological constant, as studied in sections~\ref{sec:ads} and~\ref{sec:bhs}, in the case of positive cosmological constant they are in general complex. In order to understand their relation to the Hartle-Hawking geometry given by the appropriate gluing of~\eqref{eq.dS} with~\eqref{eq.S4}, it is useful to measure the comoving ``distance''~$D$ traversed in the saddle point geometry~\eqref{eq.nbgeometry} as $r$ varies from~0 to~1: 
\begin{equation}
\label{eq.HHdistance}
D = \int_{0}^1 dr \, \sqrt{-\frac{N_{\pm}^2}{q}}.
\end{equation}
If $R_{3} \leq 1/H$, then $D$ is real and positive and one can check that the geometry~\eqref{eq.nbgeometry} is Euclidean and represents a portion of the four-sphere~\eqref{eq.S4}. In the case when $R_{3} > 1/H$, one obtains complex~$D$. The real part of~$D$ is then always equal to $\frac{\pi}{2\,H}$, which corresponds to the Euclidean distance traversed in the half-sphere part of the Hartle-Hawking geometry~\eqref{eq.S4}. The imaginary part of~$D$, which in the absence of a real part of~$D$ would correspond to a timelike separation in the Lorentzian signature, turns out to be nothing else than the proper time elapsed in the de Sitter geometry~\eqref{eq.dS} between $t = 0$ and $t = H^{-1} \, \mathrm{arcosh}{(H\, R_{3})}$, i.e. the time in which the three-sphere reaches proper size~$R_{3}$. This implies that our complex saddle-point geometry~\eqref{eq.nbgeometry} for $R_{3} > 1/H$ interpolates ``diagonally'' in the complex metric plane between the locus where~$q = R_{3}^2 > H^{-2}$ and the locus where~$q = 0$ with $q$ in between these points being a complex function of $r$. The Hartle-Hawking geometry of~\eqref{eq.S4} and~\eqref{eq.dS} achieves the same end point for $q$ by first moving in the real direction of~$D$ (the four-sphere part) and then in the imaginary direction of~$D$ (the de Sitter part), in such a way that $q$ takes real values everywhere. This geometry, which may be seen as a gluing of geometries with two different lapse values, is related to our saddle point geometry by a complex diffeomorphism and should be regarded as equivalent in our formalism\footnote{It is only in the case of a cosmological constant or adiabatic matter that the Hartle-Hawking geometry has a representation in which the scale factor is everywhere real \cite{Feldbrugge:2017fcc}. When more general matter is added, such as a scalar field in a non-constant potential, it is necessarily complex \cite{Lyons:1992ua}.}. 

The lapse action can usefully be written in the form of Eq. \eqref{eq.shiftedaction},
\begin{align}
\label{eq.dSshiftedaction}
    \frac{4G}{\pi}S_0(N) = H^4 \left(N+\frac{i}{H^2}\right)^3 - 3 (H^2R_3^2-1)\left(N+\frac{i}{H^2}\right) - \frac{2\,i}{H^2}\,.
\end{align}
At the saddle points it is complex, with value
\begin{align}
S_0(N_\pm) = \frac{\pi}{2GH^2} [- i \pm (H^2 R_3^2 -1 )^{3/2}]\,. 
\end{align}
The imaginary part determines the weighting and thus the relative probability of nucleation of a universe with this value of the cosmological constant, while the real part is associated with the classical growth of the universe up to a radius $R_3.$ This classical growth is seen as a phase in the partition function.  The fact that this phase grows fast as the universe expands, while the weighting remains constant, is an indication that the universe has become classical in a WKB sense. Note that the ``volume divergence'' is thus a welcome feature in the de Sitter case, as it is associated with the classicality of the universe. In the cosmological context we do not need, and would not want, to include counterterms. 

\begin{figure}[h]
\includegraphics[width=0.5\textwidth]{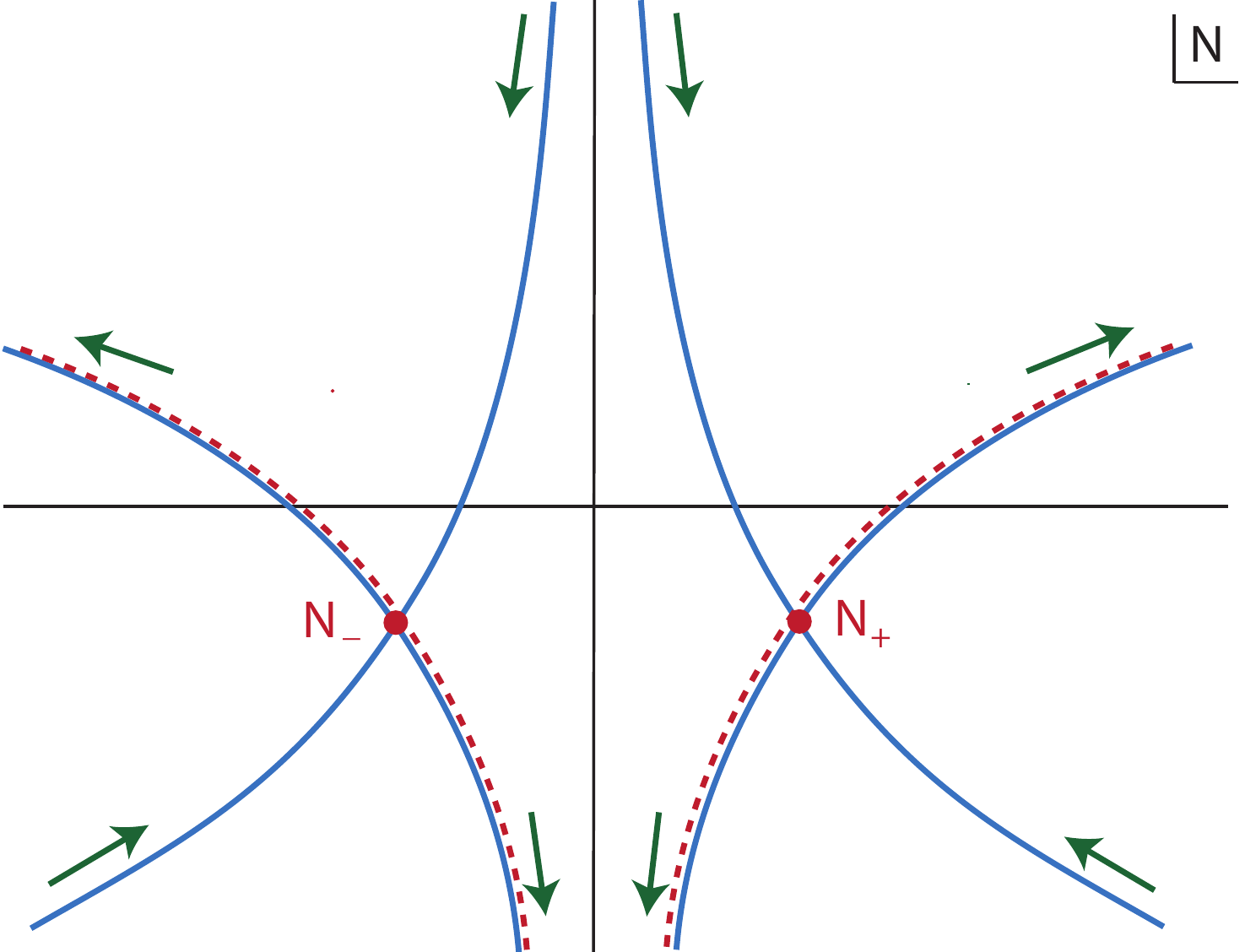}
\caption{The figure shows the structure of the flow lines with a Neumann boundary condition $\Pi_0 = - \frac{3\pi}{4G} i$ at the ``big bang'', for the case of a positive cosmological constant $\Lambda=3H^2>0$. The saddle points are complex in this case, and have equal weight. The real part of the saddle points is associated with the classical expansion of the universe while the imaginary part determines the probability of nucleation. An integration along the real $N$ line, corresponding to a Lorentzian path integral, can be deformed into the sum of the two dashed lines, passing through both saddle points. This implements the no-boundary proposal of Hartle and Hawking in the form of a minisuperspace path integral.} \label{fig:dS}
\end{figure}

The saddle points and their steepest descent flow lines are shown in Fig. \ref{fig:dS}, which should be contrasted with Fig. \ref{fig:flowAdS}. The asymptotic regions of convergence for the lapse integral are unchanged, since they are determined by the leading term in $N,$ namely $e^{iN^3/l^4}= e^{iH^4 N^3},$ cf. Eq. \eqref{eq.dSshiftedaction}. Choosing a contour running from negative imaginary infinity to the first quadrant (which we called $-{\mathcal C}_1$) would pick up only a single saddle point $N_+.$ This would yield a perfectly acceptable wavefunction/partition function, $Z \approx e^{iS(N_+)/\hbar}.$ The contour running from negative imaginary infinity to the second quadrant (which we called ${\mathcal C}_2$) would yield (minus) the complex conjugate result. By combining these two contours we have the possibility of obtaining a real wavefunction, as originally advocated by Hartle and Hawking. Just as for the AdS case, there exist two options to do so. The first is to use the sum $-({\mathcal C}_1 + {\mathcal C}_2)={\mathcal C}_0,$ which is equivalent to the Lorentzian contour. Similar arguments to those presented in section \ref{subsec:N} imply that this yields the wavefunction
\begin{align}
    Z(R_3)\mid_{{\mathcal C}_0} = e^{\frac{V_3}{4\pi G \hbar H^2}} Ai\left[\left(\frac{3V_3}{8\pi G \hbar H^2} \right)^{2/3}\left(1-H^2 R_3^2 \right) \right]\,.
\end{align}
Meanwhile, summing ${\mathcal C}_{1,2}$ such that they both run towards the upper half plane yields the result
\begin{align}
    Z(R_3)\mid_{i({\mathcal C}_2-{\mathcal C}_1)} = e^{\frac{V_3}{4\pi G \hbar H^2}} Bi\left[\left(\frac{3V_3}{8\pi G \hbar H^2} \right)^{2/3}\left(1-H^2 R_3^2 \right) \right]\,.
\end{align}
Both of these results yield an acceptable no-boundary wavefunction, the only difference being a shift in the phase, given the asymptotic expansions for real $x$
\begin{align}
    Ai(-x) \sim \cos(\frac{2}{3}x^{3/2}-\frac{\pi}{4})\,, \qquad Bi(-x) \sim \cos(\frac{2}{3}x^{3/2}+\frac{\pi}{4})\,.
\end{align}
By contrast with the negative cosmological constant case, here both options are equally viable. They both contain a trigonometric factor that one can write as the sum of two phases $\sim \left(e^{iR_3^3} + e^{-iR_3^3}\right),$ where the phases arise due to the classical expansion at late times. One can then interpret these two phases as two time-reversed universes which would decohere quickly due to the cosmological expansion \cite{Hartle:2008ng}. It is noteworthy that with positive $\Lambda$ a genuine Lorentzian path integral is viable, while with negative $\Lambda$ it was not. It will be interesting to see if this remains the case in more elaborate models, and in the presence of more general metrics\footnote{For $S^1 \times S^2$ boundary conditions and positive $\Lambda$, the saddle point geometries were already studied in \cite{Conti:2014uda}. It will be interesting to study the associated flow lines and integration contours.}.

The most important conclusion that we can draw from these observations is that the partition function in the presence of negative $\Lambda$ gives strong support to the recent implementation of the no-boundary proposal in Ref.~\cite{DiTucci:2019bui}, which used an equivalent momentum condition to \eqref{momentumcondition}. In Ref.~\cite{DiTucci:2019bui} this was called the ``no boundary term'' proposal, as the Neumann condition is obtained by not adding any surface term to the Einstein-Hilbert action. Moreover, the choice of sign in specifying the initial Euclidean expansion rate is determined in the negative $\Lambda$ case by the requirement of obtaining a sensible thermodynamic interpretation once black holes are also included. On the cosmological side this sign choice translates into picking the HH no-boundary proposal rather than Vilenkin's tunneling proposal~\cite{Vilenkin:1982de}. Expressed as a one-line conclusion, one may say that black hole thermodynamics justifies the no-boundary proposal.



\section{Discussion} \label{sec:discussion}

In this work we have provided a minisuperspace construction of gravitational partition functions in spacetimes with a negative cosmological constant, and with either $S^3$ or $S^1 \times S^2$ boundaries. Such partition functions are motivated both by classic results in black hole thermodynamics, and by the AdS/CFT correspondence. They also bear a close technical resemblance to studies in quantum cosmology, in particular in relation to the no-boundary proposal, which was another motivation for our study.

Our main findings are: 1. In the minisuperspace approach partition functions cannot be seen as sums over Euclidean metrics, but rather must  be defined as sums over certain complex classes of metrics. Despite this feature, the dominant saddle points representing AdS spacetime and AdS black holes always turn out to be Euclidean. In this way the semi-classical thermodynamic results are recovered, even though off-shell we are forced to sum over complex metrics. 2. Guided by black hole thermodynamics, we had to impose a Neumann boundary condition at the horizon of black holes in order to represent the canonical ensemble. A Dirichlet boundary condition fixing the size of the horizon would have led to a different interpretation of the partition function, more in line with calculations of the microcanonical ensemble (although this identification would require further justification). The Neumann condition, which is a condition on the expansion rate of the metric at the horizon, allows one to directly impose a regularity condition at the horizon. 3. When the outer boundary is sent to infinity, only the AdS and black hole saddle points remain of relevance. However, when the boundary resides at a finite radius, which is a scenario recently understood in terms of a dual QFT description~\cite{McGough:2016lol,Hartman:2018tkw}, three additional saddle points appear in the minisuperspace approach. Depending on parameters these subleading saddle points may be complex. As a result, providing an interpretation of these saddles\footnote{In holography, providing an interpretation of subleading saddle points in terms of dual QFT statements is challenging, yet not impossible, see Ref.~\cite{Balasubramanian:2016xho} for an example.} in the language of corresponding QFTs or ruling them out would add an element of falsifiability to the minisuperspace approach.

The fact that we had to use a Neumann boundary condition at the black hole horizons is noteworthy. In comparing with the case of a positive cosmological constant, this Neumann condition happens to be identical to the one used in a recent implementation of the no-boundary proposal \cite{DiTucci:2019bui}. There also, Dirichlet conditions proved unphysical, and a condition on the initial expansion rate of the universe was the key to obtaining a well-defined definition of the Hartle-Hawking wavefunction. As we wrote earlier, what we are finding here is that black hole thermodynamics justifies this choice, i.e. black holes thermodynamics supports the no-boundary proposal\footnote{Note that our setting is different from the ``holographic no-boundary measure'' of Hertog and Hartle \cite{Hertog:2011ky}. There they propose to use AdS/CFT on EAdS sections inside of the analytically continued saddle point geometries (which are the saddle points corresponding to a positive cosmological constant). By contrast, we start from AdS path integrals with negative cosmological constant and then simply let the cosmological constant evolve to positive values. Thus our expressions for the wave function of the universe involve a positive cosmological constant, while the ones of Hertog and Hartle involve the opposite value of the cosmological constant.}. One consequence of this is that one should no longer think of the no-boundary proposal as a sum over compact, regular metrics. Rather one should think of it as a sum over metrics with an initial Euclidean expansion rate. That the expansion rate must be Euclidean is then simply a manifestation that we are describing the quantum origin of the universe, which cannot be represented by a classical (real) solution.

Our work suggests many avenues for further study. One of them would be to understand if additional saddle points corresponding to complex geometries have any interpretation in the language of dual QFTs~\cite{McGough:2016lol}. If yes, this would provide an AdS/CFT indication about gravitational path integral including complex geometries, as we were forced to do in the minisuperspace approach. An important generalisation of our study would be to incorporate conical defects in our studies of partition functions in Section~\ref{sec:bhs}. Another very interesting direction to consider would be to generalise our minisuperspace studies away from four spacetime dimension, in which case the Neumann condition at $r = 0$ would have to be imposed differently~\cite{Krishnan:2016mcj}. More on this front, general relativity in three and especially in two spacetime dimensions does not contain dynamical gravitons and path integrals over asymptotically AdS geometries might then be well-defined. It would be very interesting to see what support such studies would give for including complexified metrics in the gravitational path integral beyond the minisuperspace approach.

\acknowledgments

ADT and JLL gratefully acknowledge the support of the European Research Council in the form of the ERC Consolidator Grant CoG 772295 ``Qosmology''. MPH and the Gravity, Quantum Fields and Information
group at AEI are supported by the Alexander
von Humboldt Foundation and the Federal Ministry for
Education and Research through the Sofja Kovalevskaja
Award.

\appendix

\section{Fluctuation determinant for mixed Neumann-Dirichlet boundary conditions} \label{sec:determinant}

In evaluating our path integrals, we could make use of the fact that the actions were quadratic in the scale factors, thus allowing a decomposition of the path into a classical solution $\bar{q}$ and a fluctuation $Q$, i.e. $q(r) = \bar{q}(r) + Q(r),$ with the resulting path integral over $Q$ being of Gaussian form,
\begin{align}
F(N) = \int_{\dot{Q}(0)=0}^{Q(1)=0} D[Q] e^{i\int_0^1 dr \frac{\dot{Q}^2}{N}}\,,    
\end{align}
where we have neglected an unimportant numerical factor in the exponent. To ensure that the total scale factor $q$ satisfies the mixed Neumann-Dirichlet boundary conditions, the fluctuation must satisfy $\dot{Q}(0)=0$ and $Q(1)=0.$ Here we would like to determine the dependence of the above integral on the lapse $N.$ To do so, we will use a re-scaled coordinate $\tilde{r}=rN,$ with range $0 \leq \tilde{r} \leq N.$ The integral then becomes 
\begin{align}
F(N) &= \int_{Q_{,\tilde{r}}(0)=0}^{Q(1)=0} D[Q] e^{i\int_0^N d\tilde{r} {Q}_{,\tilde{r}}^2} \nonumber \\ &= \int_{Q_{,\tilde{r}}(0)=0}^{Q(1)=0} D[Q] e^{- i\int_0^N {Q}_{,\tilde{r}} \frac{d^2}{d\tilde{r}^2} {Q}_{,\tilde{r}}} = \sqrt{\frac{2}{\pi i}}\left[\text{det}\left(-\frac{d^2}{d\tilde{r}^2}\right)\right]^{-1/2}\,.
\end{align}
With the assumed boundary conditions, the operator $-\frac{d^2}{d\tilde{r}^2}$ satisfies the eigenvalue equation $-\frac{d^2}{d\tilde{r}^2} x_n = \lambda_n x_n$ with eigenfunctions $x_n$ and eigenvalues $\lambda_n,$
\begin{align}
    x_n = a_n \cos\left[\frac{(2n+1)\pi}{2N}\tilde{r} \right]\,, \quad \lambda_n = \left[\frac{(2n+1)\pi}{2N}\right]^2\,, \quad n \in \mathbb{N}\,.
\end{align}
The determinant is given by the product of all eigenvalues. We can evaluate it using zeta function regularisation (see e.g. Ref.~\cite{Grosche:1998yu}). Thus in analogy with the zeta function $\zeta(s)=\sum_{n \in \mathbb{N}}n^{-s}$ we define
\begin{align}
    \zeta_\lambda(s) \equiv \sum_{n \in \mathbb{N}} \lambda_n^{-s} = \left(\frac{2N}{\pi} \right)^{2s}\sum_{n\in \mathbb{N}} \frac{1}{(2n+1)^{2s}}\,.
\end{align}
The last term corresponds to the zeta function where one would sum only over odd terms. We can obtain this sum by subtracting the even terms,
\begin{align}
    1+\frac{1}{3^{2s}}+\frac{1}{5^{2s}}+\cdots & = 1+\frac{1}{2^{2s}}+\frac{1}{3^{2s}}+\cdots - [\frac{1}{2^{2s}}+\frac{1}{4^{2s}}+\cdots] \nonumber \\ & = 1+\frac{1}{2^{2s}}+\frac{1}{3^{2s}}+\cdots - \frac{1}{2^{2s}} [1+ \frac{1}{2^{2s}}+\frac{1}{3^{2s}}+\cdots]\,.
\end{align}
Hence we obtain 
\begin{align}
    \zeta_\lambda(s) = \left(\frac{2N}{\pi} \right)^{2s} \left( 1 - 2^{-2s}\right)\zeta(s)\,.
\end{align}
The zeta function can be analytically continued to $s=0,$ where the derivative $\zeta_\lambda^\prime(0)$ is related to the product of all $\lambda_n$ such that
\begin{align}
    \left[\text{det}\left(-\frac{d^2}{d\tilde{r}^2}\right)\right] = e^{-\zeta_\lambda^\prime(0)}=2\,,
\end{align}
where we have made use of $\zeta(0)=-\frac{1}{2}.$ In the end we find the remarkably simple result that 
\begin{align}
    F(N) = \frac{1}{\sqrt{\pi \, i}}\,.
\end{align}
In particular, note that the fluctuation determinant for the Neumann-Dirichlet problem does not contain any dependence on the lapse $N,$ unlike in the well known pure Dirichlet case where the determinant is proportional to $N^{-1/2}$ \cite{Grosche:1998yu}.

\bibliographystyle{utphys}
\bibliography{AdSPathIntegralBibliography}

\providecommand{\href}[2]{#2}\begingroup\raggedright\begin{thebibliography}{10}

\bibitem{Hartle:1976tp}
J.~Hartle and S.~Hawking, ``{Path Integral Derivation of Black Hole
  Radiance},'' \href{http://dx.doi.org/10.1103/PhysRevD.13.2188}{{\em Phys.
  Rev. D} {\bfseries 13} (1976) 2188--2203}.

\bibitem{Hartle:1983ai}
J.~B. Hartle and S.~W. Hawking, ``Wave function of the universe,'' {\em Phys.
  Rev. D} {\bfseries 28} (1983) 2960--2975.

\bibitem{Hawking:1982dh}
S.~Hawking and D.~N. Page, ``{Thermodynamics of Black Holes in anti-De Sitter
  Space},'' \href{http://dx.doi.org/10.1007/BF01208266}{{\em Commun. Math.
  Phys.} {\bfseries 87} (1983) 577}.

\bibitem{Maldacena:1997re}
J.~M. Maldacena, ``{The Large N limit of superconformal field theories and
  supergravity},'' \href{http://dx.doi.org/10.1023/A:1026654312961}{{\em Int.
  J. Theor. Phys.} {\bfseries 38} (1999) 1113--1133},
  \href{http://arxiv.org/abs/hep-th/9711200}{{\ttfamily arXiv:hep-th/9711200}}.

\bibitem{Gubser:1998bc}
S.~Gubser, I.~R. Klebanov, and A.~M. Polyakov, ``{Gauge theory correlators from
  noncritical string theory},''
  \href{http://dx.doi.org/10.1016/S0370-2693(98)00377-3}{{\em Phys. Lett. B}
  {\bfseries 428} (1998) 105--114},
  \href{http://arxiv.org/abs/hep-th/9802109}{{\ttfamily arXiv:hep-th/9802109}}.

\bibitem{Witten:1998qj}
E.~Witten, ``{Anti-de Sitter space and holography},''
  \href{http://dx.doi.org/10.4310/ATMP.1998.v2.n2.a2}{{\em Adv. Theor. Math.
  Phys.} {\bfseries 2} (1998) 253--291},
  \href{http://arxiv.org/abs/hep-th/9802150}{{\ttfamily arXiv:hep-th/9802150}}.

\bibitem{Witten:1998zw}
E.~Witten, ``{Anti-de Sitter space, thermal phase transition, and confinement
  in gauge theories},''
  \href{http://dx.doi.org/10.4310/ATMP.1998.v2.n3.a3}{{\em Adv. Theor. Math.
  Phys.} {\bfseries 2} (1998) 505--532},
  \href{http://arxiv.org/abs/hep-th/9803131}{{\ttfamily arXiv:hep-th/9803131}}.

\bibitem{Halliwell:1988wc}
J.~J. Halliwell, ``{Derivation of the Wheeler-De Witt Equation from a Path
  Integral for Minisuperspace Models},''
  \href{http://dx.doi.org/10.1103/PhysRevD.38.2468}{{\em Phys. Rev. D}
  {\bfseries 38} (1988) 2468}.

\bibitem{DiTucci:2019bui}
A.~Di~Tucci, J.-L. Lehners, and L.~Sberna, ``{No-boundary prescriptions in
  Lorentzian quantum cosmology},''
  \href{http://dx.doi.org/10.1103/PhysRevD.100.123543}{{\em Phys. Rev. D}
  {\bfseries 100} no.~12, (2019) 123543},
  \href{http://arxiv.org/abs/1911.06701}{{\ttfamily arXiv:1911.06701
  [hep-th]}}.

\bibitem{Feldbrugge:2017kzv}
J.~Feldbrugge, J.-L. Lehners, and N.~Turok, ``{Lorentzian Quantum Cosmology},''
  \href{http://dx.doi.org/10.1103/PhysRevD.95.103508}{{\em Phys. Rev. D}
  {\bfseries 95} no.~10, (2017) 103508},
  \href{http://arxiv.org/abs/1703.02076}{{\ttfamily arXiv:1703.02076
  [hep-th]}}.

\bibitem{Feldbrugge:2017fcc}
J.~Feldbrugge, J.-L. Lehners, and N.~Turok, ``{No smooth beginning for
  spacetime},'' \href{http://dx.doi.org/10.1103/PhysRevLett.119.171301}{{\em
  Phys. Rev. Lett.} {\bfseries 119} no.~17, (2017) 171301},
  \href{http://arxiv.org/abs/1705.00192}{{\ttfamily arXiv:1705.00192
  [hep-th]}}.

\bibitem{Feldbrugge:2017mbc}
J.~Feldbrugge, J.-L. Lehners, and N.~Turok, ``{No rescue for the no boundary
  proposal: Pointers to the future of quantum cosmology},''
  \href{http://dx.doi.org/10.1103/PhysRevD.97.023509}{{\em Phys. Rev. D}
  {\bfseries 97} no.~2, (2018) 023509},
  \href{http://arxiv.org/abs/1708.05104}{{\ttfamily arXiv:1708.05104
  [hep-th]}}.

\bibitem{DiazDorronsoro:2017hti}
J.~Diaz~Dorronsoro, J.~J. Halliwell, J.~B. Hartle, T.~Hertog, and O.~Janssen,
  ``{Real no-boundary wave function in Lorentzian quantum cosmology},''
  \href{http://dx.doi.org/10.1103/PhysRevD.96.043505}{{\em Phys. Rev. D}
  {\bfseries 96} no.~4, (2017) 043505},
  \href{http://arxiv.org/abs/1705.05340}{{\ttfamily arXiv:1705.05340 [gr-qc]}}.

\bibitem{DiazDorronsoro:2018wro}
J.~Diaz~Dorronsoro, J.~J. Halliwell, J.~B. Hartle, T.~Hertog, O.~Janssen, and
  Y.~Vreys, ``{Damped perturbations in the no-boundary state},''
  \href{http://dx.doi.org/10.1103/PhysRevLett.121.081302}{{\em Phys. Rev.
  Lett.} {\bfseries 121} no.~8, (2018) 081302},
  \href{http://arxiv.org/abs/1804.01102}{{\ttfamily arXiv:1804.01102 [gr-qc]}}.

\bibitem{Vilenkin:2018dch}
A.~Vilenkin and M.~Yamada, ``{Tunneling wave function of the universe},''
  \href{http://dx.doi.org/10.1103/PhysRevD.98.066003}{{\em Phys. Rev. D}
  {\bfseries 98} no.~6, (2018) 066003},
  \href{http://arxiv.org/abs/1808.02032}{{\ttfamily arXiv:1808.02032 [gr-qc]}}.

\bibitem{Vilenkin:2018oja}
A.~Vilenkin and M.~Yamada, ``{Tunneling wave function of the universe II: the
  backreaction problem},''
  \href{http://dx.doi.org/10.1103/PhysRevD.99.066010}{{\em Phys. Rev. D}
  {\bfseries 99} no.~6, (2019) 066010},
  \href{http://arxiv.org/abs/1812.08084}{{\ttfamily arXiv:1812.08084 [gr-qc]}}.

\bibitem{Feldbrugge:2018gin}
J.~Feldbrugge, J.-L. Lehners, and N.~Turok, ``{Inconsistencies of the New
  No-Boundary Proposal},''
  \href{http://dx.doi.org/10.3390/universe4100100}{{\em Universe} {\bfseries 4}
  no.~10, (2018) 100}, \href{http://arxiv.org/abs/1805.01609}{{\ttfamily
  arXiv:1805.01609 [hep-th]}}.

\bibitem{DiTucci:2019dji}
A.~Di~Tucci and J.-L. Lehners, ``{No-Boundary Proposal as a Path Integral with
  Robin Boundary Conditions},''
  \href{http://dx.doi.org/10.1103/PhysRevLett.122.201302}{{\em Phys. Rev.
  Lett.} {\bfseries 122} no.~20, (2019) 201302},
  \href{http://arxiv.org/abs/1903.06757}{{\ttfamily arXiv:1903.06757
  [hep-th]}}.

\bibitem{Louko:1988bk}
J.~Louko, ``{Canonizing the Hartle-hawking Proposal},''
  \href{http://dx.doi.org/10.1016/0370-2693(88)90008-1}{{\em Phys. Lett. B}
  {\bfseries 202} (1988) 201--206}.

\bibitem{Halliwell:1988ik}
J.~J. Halliwell and J.~Louko, ``{Steepest Descent Contours in the Path Integral
  Approach to Quantum Cosmology. 1. The De Sitter Minisuperspace Model},''
  \href{http://dx.doi.org/10.1103/PhysRevD.39.2206}{{\em Phys. Rev. D}
  {\bfseries 39} (1989) 2206}.

\bibitem{Gibbons:1976ue}
G.~Gibbons and S.~Hawking, ``{Action Integrals and Partition Functions in
  Quantum Gravity},'' \href{http://dx.doi.org/10.1103/PhysRevD.15.2752}{{\em
  Phys. Rev. D} {\bfseries 15} (1977) 2752--2756}.

\bibitem{Witten:2010cx}
E.~Witten, ``{Analytic Continuation Of Chern-Simons Theory},'' {\em AMS/IP
  Stud. Adv. Math.} {\bfseries 50} (2011) 347--446,
  \href{http://arxiv.org/abs/1001.2933}{{\ttfamily arXiv:1001.2933 [hep-th]}}.

\bibitem{McGough:2016lol}
L.~McGough, M.~Mezei, and H.~Verlinde, ``{Moving the CFT into the bulk with $
  T\overline{T} $},'' \href{http://dx.doi.org/10.1007/JHEP04(2018)010}{{\em
  JHEP} {\bfseries 04} (2018) 010},
  \href{http://arxiv.org/abs/1611.03470}{{\ttfamily arXiv:1611.03470
  [hep-th]}}.

\bibitem{Hartman:2018tkw}
T.~Hartman, J.~Kruthoff, E.~Shaghoulian, and A.~Tajdini, ``{Holography at
  finite cutoff with a $T^2$ deformation},''
  \href{http://dx.doi.org/10.1007/JHEP03(2019)004}{{\em JHEP} {\bfseries 03}
  (2019) 004}, \href{http://arxiv.org/abs/1807.11401}{{\ttfamily
  arXiv:1807.11401 [hep-th]}}.

\bibitem{Jafferis:2011zi}
D.~L. Jafferis, I.~R. Klebanov, S.~S. Pufu, and B.~R. Safdi, ``{Towards the
  F-Theorem: N=2 Field Theories on the Three-Sphere},''
  \href{http://dx.doi.org/10.1007/JHEP06(2011)102}{{\em JHEP} {\bfseries 06}
  (2011) 102}, \href{http://arxiv.org/abs/1103.1181}{{\ttfamily arXiv:1103.1181
  [hep-th]}}.

\bibitem{Carter:1973rla}
B.~Carter, ``{Black holes equilibrium states},'' in {\em {Les Houches Summer
  School of Theoretical Physics}: {Black Holes}}, pp.~57--214.
\newblock 1973.

\bibitem{Balasubramanian:1999re}
V.~Balasubramanian and P.~Kraus, ``{A Stress tensor for Anti-de Sitter
  gravity},'' \href{http://dx.doi.org/10.1007/s002200050764}{{\em Commun. Math.
  Phys.} {\bfseries 208} (1999) 413--428},
  \href{http://arxiv.org/abs/hep-th/9902121}{{\ttfamily arXiv:hep-th/9902121}}.

\bibitem{Halliwell:1990tu}
J.~J. Halliwell and J.~Louko, ``{Steepest Descent Contours in the Path Integral
  Approach to Quantum Cosmology. 3. A General Method With Applications to
  Anisotropic Minisuperspace Models},''
  \href{http://dx.doi.org/10.1103/PhysRevD.42.3997}{{\em Phys. Rev. D}
  {\bfseries 42} (1990) 3997--4031}.

\bibitem{Caputa:2018asc}
P.~Caputa and S.~Hirano, ``{Airy Function and 4d Quantum Gravity},''
  \href{http://dx.doi.org/10.1007/JHEP06(2018)106}{{\em JHEP} {\bfseries 06}
  (2018) 106}, \href{http://arxiv.org/abs/1804.00942}{{\ttfamily
  arXiv:1804.00942 [hep-th]}}.

\bibitem{Donnelly:2019pie}
W.~Donnelly, E.~LePage, Y.-Y. Li, A.~Pereira, and V.~Shyam, ``{Quantum
  corrections to finite radius holography and holographic entanglement
  entropy},'' \href{http://dx.doi.org/10.1007/JHEP05(2020)006}{{\em JHEP}
  {\bfseries 05} (2020) 006}, \href{http://arxiv.org/abs/1909.11402}{{\ttfamily
  arXiv:1909.11402 [hep-th]}}.

\bibitem{Hirano:2019szi}
S.~Hirano, ``{Quantum Holographic Entanglement Entropy to All Orders in $1/N$
  Expansion},'' \href{http://dx.doi.org/10.1093/ptep/ptaa019}{{\em PTEP}
  {\bfseries 2020} no.~4, (2020) 043B02},
  \href{http://arxiv.org/abs/1911.01640}{{\ttfamily arXiv:1911.01640
  [hep-th]}}.

\bibitem{York:1972sj}
J.~York, James~W., ``{Role of conformal three geometry in the dynamics of
  gravitation},'' \href{http://dx.doi.org/10.1103/PhysRevLett.28.1082}{{\em
  Phys. Rev. Lett.} {\bfseries 28} (1972) 1082--1085}.

\bibitem{Teitelboim:1981ua}
C.~Teitelboim, ``{Quantum Mechanics of the Gravitational Field},''
  \href{http://dx.doi.org/10.1103/PhysRevD.25.3159}{{\em Phys. Rev. D}
  {\bfseries 25} (1982) 3159}.

\bibitem{Vallee}
O.~Vallee and M.~Soares, {\em {Airy functions and applications to physics}}.
\newblock Imperial College Press, 2004.

\bibitem{deHaro:2000vlm}
S.~de~Haro, S.~N. Solodukhin, and K.~Skenderis, ``{Holographic reconstruction
  of space-time and renormalization in the AdS / CFT correspondence},''
  \href{http://dx.doi.org/10.1007/s002200100381}{{\em Commun. Math. Phys.}
  {\bfseries 217} (2001) 595--622},
  \href{http://arxiv.org/abs/hep-th/0002230}{{\ttfamily arXiv:hep-th/0002230}}.

\bibitem{Aharony:2008ug}
O.~Aharony, O.~Bergman, D.~L. Jafferis, and J.~Maldacena, ``{N=6 superconformal
  Chern-Simons-matter theories, M2-branes and their gravity duals},''
  \href{http://dx.doi.org/10.1088/1126-6708/2008/10/091}{{\em JHEP} {\bfseries
  10} (2008) 091}, \href{http://arxiv.org/abs/0806.1218}{{\ttfamily
  arXiv:0806.1218 [hep-th]}}.

\bibitem{Pestun:2007rz}
V.~Pestun, ``{Localization of gauge theory on a four-sphere and supersymmetric
  Wilson loops},'' \href{http://dx.doi.org/10.1007/s00220-012-1485-0}{{\em
  Commun. Math. Phys.} {\bfseries 313} (2012) 71--129},
  \href{http://arxiv.org/abs/0712.2824}{{\ttfamily arXiv:0712.2824 [hep-th]}}.

\bibitem{Fuji:2011km}
H.~Fuji, S.~Hirano, and S.~Moriyama, ``{Summing Up All Genus Free Energy of
  ABJM Matrix Model},'' \href{http://dx.doi.org/10.1007/JHEP08(2011)001}{{\em
  JHEP} {\bfseries 08} (2011) 001},
  \href{http://arxiv.org/abs/1106.4631}{{\ttfamily arXiv:1106.4631 [hep-th]}}.

\bibitem{Marino:2011eh}
M.~Marino and P.~Putrov, ``{ABJM theory as a Fermi gas},''
  \href{http://dx.doi.org/10.1088/1742-5468/2012/03/P03001}{{\em J. Stat.
  Mech.} {\bfseries 1203} (2012) P03001},
  \href{http://arxiv.org/abs/1110.4066}{{\ttfamily arXiv:1110.4066 [hep-th]}}.

\bibitem{Krishnan:2016mcj}
C.~Krishnan and A.~Raju, ``{A Neumann Boundary Term for Gravity},''
  \href{http://dx.doi.org/10.1142/S0217732317500778}{{\em Mod. Phys. Lett. A}
  {\bfseries 32} no.~14, (2017) 1750077},
  \href{http://arxiv.org/abs/1605.01603}{{\ttfamily arXiv:1605.01603
  [hep-th]}}.

\bibitem{Brown:1994gs}
J.~Brown, J.~Creighton, and R.~B. Mann, ``{Temperature, energy and heat
  capacity of asymptotically anti-de Sitter black holes},''
  \href{http://dx.doi.org/10.1103/PhysRevD.50.6394}{{\em Phys. Rev. D}
  {\bfseries 50} (1994) 6394--6403},
  \href{http://arxiv.org/abs/gr-qc/9405007}{{\ttfamily arXiv:gr-qc/9405007}}.

\bibitem{deBoer:2015ija}
J.~de~Boer, M.~P. Heller, and N.~Pinzani-Fokeeva, ``{Effective actions for
  relativistic fluids from holography},''
  \href{http://dx.doi.org/10.1007/JHEP08(2015)086}{{\em JHEP} {\bfseries 08}
  (2015) 086}, \href{http://arxiv.org/abs/1504.07616}{{\ttfamily
  arXiv:1504.07616 [hep-th]}}.

\bibitem{Brown:1992bq}
J.~Brown and J.~York, James~W., ``{The Microcanonical functional integral. 1.
  The Gravitational field},''
  \href{http://dx.doi.org/10.1103/PhysRevD.47.1420}{{\em Phys. Rev. D}
  {\bfseries 47} (1993) 1420--1431},
  \href{http://arxiv.org/abs/gr-qc/9209014}{{\ttfamily arXiv:gr-qc/9209014}}.

\bibitem{Halliwell:1984eu}
J.~Halliwell and S.~Hawking, ``{The Origin of Structure in the Universe},''
  \href{http://dx.doi.org/10.1103/PhysRevD.31.1777}{{\em Adv. Ser. Astrophys.
  Cosmol.} {\bfseries 3} (1987) 277--291}.

\bibitem{Halliwell:1989dy}
J.~J. Halliwell and J.~B. Hartle, ``{Integration Contours for the No Boundary
  Wave Function of the Universe},''
  \href{http://dx.doi.org/10.1103/PhysRevD.41.1815}{{\em Phys. Rev. D}
  {\bfseries 41} (1990) 1815}.

\bibitem{Lyons:1992ua}
G.~Lyons, ``{Complex solutions for the scalar field model of the universe},''
  \href{http://dx.doi.org/10.1103/PhysRevD.46.1546}{{\em Phys. Rev. D}
  {\bfseries 46} (1992) 1546--1550}.

\bibitem{Hartle:2008ng}
J.~B. Hartle, S.~Hawking, and T.~Hertog, ``{The Classical Universes of the
  No-Boundary Quantum State},''
  \href{http://dx.doi.org/10.1103/PhysRevD.77.123537}{{\em Phys. Rev. D}
  {\bfseries 77} (2008) 123537},
  \href{http://arxiv.org/abs/0803.1663}{{\ttfamily arXiv:0803.1663 [hep-th]}}.

\bibitem{Conti:2014uda}
G.~Conti and T.~Hertog, ``{Two wave functions and dS/CFT on S$^{1}$ x
  S$^{2}$},'' \href{http://dx.doi.org/10.1007/JHEP06(2015)101}{{\em JHEP}
  {\bfseries 06} (2015) 101}, \href{http://arxiv.org/abs/1412.3728}{{\ttfamily
  arXiv:1412.3728 [hep-th]}}.

\bibitem{Vilenkin:1982de}
A.~Vilenkin, ``{Creation of Universes from Nothing},''
  \href{http://dx.doi.org/10.1016/0370-2693(82)90866-8}{{\em Phys. Lett. B}
  {\bfseries 117} (1982) 25--28}.

\bibitem{Balasubramanian:2016xho}
V.~Balasubramanian, A.~Bernamonti, B.~Craps, T.~De~Jonckheere, and F.~Galli,
  ``{Entwinement in discretely gauged theories},''
  \href{http://dx.doi.org/10.1007/JHEP12(2016)094}{{\em JHEP} {\bfseries 12}
  (2016) 094}, \href{http://arxiv.org/abs/1609.03991}{{\ttfamily
  arXiv:1609.03991 [hep-th]}}.

\bibitem{Hertog:2011ky}
T.~Hertog and J.~Hartle, ``{Holographic No-Boundary Measure},''
  \href{http://dx.doi.org/10.1007/JHEP05(2012)095}{{\em JHEP} {\bfseries 05}
  (2012) 095}, \href{http://arxiv.org/abs/1111.6090}{{\ttfamily arXiv:1111.6090
  [hep-th]}}.

\bibitem{Grosche:1998yu}
C.~Grosche and F.~Steiner, {\em {Handbook of Feynman Path Integrals}},
  vol.~145.
\newblock 1998.

\end{thebibliography}\endgroup

\end{document}